\documentclass{mn2e}

\newcommand{\etal}{{et al.}}
\newcommand{\kms}{\mbox{${\,\rm km~s}^{-1}$}\,}
\newcommand{\kev}{\,\mbox{${\,\rm keV}$\,}}
\newcommand{\ergs}{\mbox{${\,\rm erg~s}^{-1}$\,}}
\newcommand{\msunyr}{\mbox{\,{\rm M}$_\odot$\ {\rm yr}$^{-1}$\,}}
\newcommand{\msun}{\mbox{\,{\rm M}$_\odot$\,}}

\title[Chandra \& XMM-Newton Observations of NGC\,5253]{Chandra \&
XMM-Newton Observations of NGC\,5253. Analysis of the X-ray Emission from a
Dwarf Starburst Galaxy.}

\author[L. K. Summers \etal]{Lesley K. Summers$^{1}$,
Ian R. Stevens$^{1}$, David K. Strickland$^{2}$\thanks{{\it Chandra}
Fellow} \newauthor and
Timothy M. Heckman$^{2}$\\
$^{1}$ School of Physics \& Astronomy, University of Birmingham,
Edgbaston, Birmingham, B15 2TT, UK\\
lks@star.sr.bham.ac.uk; irs@star.sr.bham.ac.uk\\
$^{2}$ Department of Physics \& Astronomy, The Johns Hopkins University,
3400 North Charles Street, Baltimore, MD 21218, USA\\
dks@pha.jhu.edu; heckman@pha.jhu.edu\\}

\date{Accepted .....................; Received .....................; 
in original form .......................}

\begin{document}

\maketitle

\begin{abstract} 

We present {\it Chandra} and {\it XMM-Newton} X-ray data of NGC\,5253, a
local starbursting dwarf elliptical galaxy, in the early stages of a
starburst episode. Contributions to the X-ray emission come from
discrete point sources and extended diffuse emission, in the form of
what appear to be multiple superbubbles, and smaller bubbles probably
associated with individual star clusters. {\it Chandra} detects 17
sources within the optical extent of NGC\,5253 down to a completeness
level corresponding to a luminosity of $1.5 \times 10^{37}\ergs$. The
slope of the point source X-ray luminosity function is $-0.54
\pm^{0.21}_{0.16}$, similar to that of other nearby dwarf starburst
galaxies. Several different types of source are detected within the
galaxy, including X-ray binaries and the emission associated with
star-clusters.

Comparison of the diffuse X-ray emission with the observed H$\alpha$
emission shows similarities in their extent. The best spectral fit to
the diffuse emission is obtained with an absorbed, two temperature model
giving temperatures for the two gas components of $\sim 0.24\kev$ and
$\sim 0.75\kev$. The derived parameters of the diffuse X-ray emitting
gas are as follows: a total mass of $\sim 1.4 \times 10^{6}
f^{1/2}\msun$, where $f$ is the volume filling factor of the X-ray
emitting gas, and a total thermal energy content for the hot X-ray
emitting gas of $\sim 3.4 \times 10^{54}f^{1/2}$~erg. The pressure in
the diffuse gas is $P/k\sim 10^{6} f^{-1/2}$~K~cm$^{-3}$.  We find that
these values are broadly commensurate with the mass and energy injection
from the starburst population.  Analysis of the kinematics of the
starburst region suggest that the stellar ejecta contained within it can
escape the gravitational potential well of the galaxy, and pollute the
surrounding IGM.

\end{abstract}


\section{Introduction}

NGC\,5253 is a starbursting dwarf elliptical galaxy which lies at a
distance of $\sim 3.15$~Mpc (Freedman \etal\ 2001) and with an
inclination of 67$^{\circ}$ (the mean value quoted by the Lyon/Meudon
Extra-galactic Database [LEDA]). The signature of Wolf-Rayet stars have
been detected in the nucleus of the galaxy (Schaerer \etal\ 1997, Walsh
\& Roy 1987) implying that this is a young starburst and as such allows
an opportunity for the study of the starburst phenomenon in its earlier
stages of development and its effect on galaxy evolution in the local
Universe. In fact, Rieke, Lebofsky \& Walker (1988) went so far as to
classify it as one of the youngest starbursts known. Dwarf galaxies, as
the basic building blocks in the hierarchical merging cosmology
scenario, are likely to have harboured the earliest sites of
star-formation in the Universe and so their study in the local Universe
can give insight into the evolution of such objects at high
redshift. Observations of local edge-on starburst galaxies (Strickland
\etal\ 2000; Weaver 2001) are presenting a picture of kpc-scale, soft
X-ray emitting, bipolar outflows in the form of galactic winds
transporting mass, newly synthesised heavy elements and energy into the
intergalactic medium (IGM). A similar outflow is seen in the dwarf
starburst galaxy NGC\,1569 (Martin, Kobulnicky \& Heckman 2002), while
the dwarf NGC\,4449 shows what may be the beginnings of a galactic wind
emerging from an extended superbubble (Summers \etal\ 2003). The same
situation seems to be the case in the NGC\,3077 dwarf (Ott, Martin \&
Walter 2003). These winds result from the pressure driven outflows along
these galaxy's minor axis produced from the efficient thermalization of
the mechanical energy from the supernovae (SN) explosions and stellar
winds of the massive stars in their OB associations and super
star-clusters (SSC).  NGC\,5253 being both inclined to our line-of-sight
and young presents a less clearly observable picture. Its youth means
that it is unlikely to have developed a superwind and the diffuse X-ray
emission observed is much more likely to be associated with multiple
superbubbles around its OB associations and SSC, as suggested from
earlier {\it ROSAT} observations of this galaxy (Strickland \& Stevens
1999).

Observations of NGC\,5253 at radio wavelengths have shown it to have a
very flat centimetre-wavelength continuum which is indicative of thermal
emission from H{\small II} regions, with only low levels of synchrotron
emission from supernova remnants (Beck \etal\ 1996; Turner, Ho \& Beck 1998).
Ferrarese \etal\ (2000) quote a low metallicity of $0.2 Z_{\odot}$,
based on the work of Webster \& Smith (1983).
This figure suggests that a large amount of metal enrichment
has not occurred in the central H{\small II} regions, and may be further
evidence of the youth of the starburst region. The total and H{\small I} masses
for the galaxy are calculated to be $6.4 \times
10^{8}\msun$ and $8.3 \times 10^{7}\msun$ from 21~cm observations (Reif
\etal\ 1982), corrected for the
difference in assumed distance to NGC\,5253. Later VLA 21~cm
observations show NGC\,5253 to be peculiar, in so much as its neutral
hydrogen appears to rotate about the optical major axis of the galaxy
(Kobulnicky \& Skillman 1995). However, CO observations (Turner, Beck \& 
Hurt 1997; Meier, Turner \& Beck 2002) detect sources which are coincident
with the dust lane seen to the SE of the nucleus of the galaxy in
optical images, and these CO clouds appear to be infalling into
NGC\,5253. Consequently, the dynamical situation in NGC\,5253, rotation
or infall, is not completely clear.

Infrared emission is detected from what seems to be a highly obscured,
massive ($10^{5} -10^{6}$ stars), small ($\sim$ a few pc diameter) and
very young globular cluster in the central starburst region of the
galaxy (Gorjian, Turner \& Beck 2001).  This source contributes $\sim
50\%$ of the total observed infrared luminosity of the galaxy.  A
summary of some of the properties of NGC\,5253 is given in Table~1,
along with estimates of the current star-formation rate of the galaxy
estimated from far-infrared and H$\alpha$ fluxes.

Caldwell \& Phillips (1989), using broadband multicolour (UBV) CCD
images, narrow-band H$\alpha$ and long-slit Ca{\small II} triplet
spectra, determined that NGC\,5253 had undergone an increased rate of
star-formation throughout the galaxy about $10^{8} - 10^{9}$~yr ago, as
suggested by the presence of over 100 star clusters of a similar age
lying outside the nuclear region. They also established that the intense
star-formation now occurring in the nuclear region of the galaxy began
around 10~Myr ago, giving an upper limit for the age of superbubbles
associated with it.  More recently, Tremonti et al. (2001) analysed
several clusters in NGC\,5253, determining ages in the range of 1 to
8~Myr. Calzetti et al. (1997) also noted two other clusters with older
ages, determined to be between 10 and 50~Myr.

The two very bright SN explosions seen in this galaxy in 1895 and 1972
are amongst the apparently brightest recorded extra-galactic SNe
(Ardeberg \& de Groot 1973) and may be associated with the large scale
star-formation, although neither was particularly centrally located. The
Fabry-Perot H$\alpha$ images of Marlowe \etal\ (1995) show that the
ionized gas in the galaxy has a complex distribution, consisting of a
bright central region extending $\sim 30{''}$ from the centre in a N-S
direction, embedded within a system of loops and filaments, with an
overall extent of $\sim 2{'}$. To the NW and WSW of the centre are 2
large superbubbles with diameters $\sim 1{'}$, expanding at $\sim
35\kms$, while to the east side of the galaxy, a pair of radially
oriented filaments are detected. This latter observation is suggestive
of the rupture of a superbubble blown by the stellar activity in the
central regions.

NGC\,5253 has previously been observed at X-ray wavelengths with the
{\it Einstein} IPC (Fabbiano, Kim \& Trinchieri 1992) and both the {\it
ROSAT} PSPC (Martin \& Kennicutt 1995; Stevens \& Strickland 1998) and
HRI (Strickland \& Stevens 1999). The {\it Einstein} IPC and {\it ROSAT}
PSPC results showed soft thermal X-ray emission from what appeared to be
an extended source, suggestive of a luminous superbubble. The better
spatial resolution of the {\it ROSAT} HRI instrument detected a complex
of at least five X-ray sources, that could be young superbubbles blown
by individual young star clusters in the starburst region. The total
X-ray luminosity determined from these observations was similar and lies
in the range of $(2.4 - 4.1) \times 10^{38}\ergs$, depending on the value
adopted for the absorbing column density.

In Section 2 we describe the {\it Chandra} and {\it XMM-Newton}
observations. The X-ray emission from the whole galaxy, the point sources
and the diffuse emission are discussed in Section~3. Section~4 contains a
more general discussion of the morphology of the X-ray emission and its
relationship to emission from other wavebands, along with a determination
of the dynamics of the expanding superbubbles of the central region and
their effect on NGC\,5253. Our main conclusions are summarised in Section~5.

\begin{figure*}
\vspace{13cm} 
\includegraphics{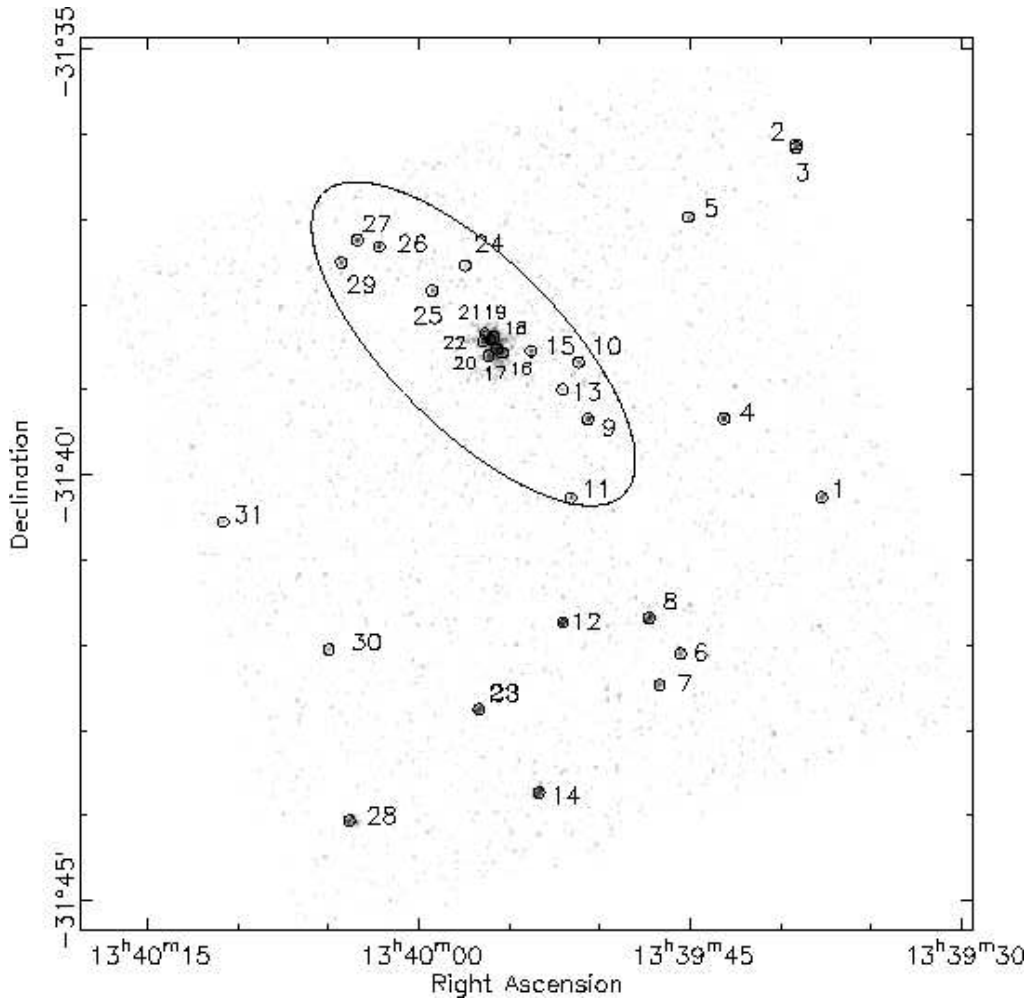}
\caption{Low resolution (smoothed using a Gaussian with FWHM of 4 pixels
$\sim 2^{''}$), background subtracted {\it Chandra} image, in
the $0.3 - 8.0\kev$ energy band, showing the full ACIS-S3 chip field of
view, marked with the 31 point sources (Table~2). The
image is an inverted log greyscale with the flux density ranging from
$7.63 \times 10^{-14}\ergs$~cm$^{-2}$~arcmin$^{-2}$ to $9.72 \times
10^{-11}\ergs$~cm$^{-2}$~arcmin$^{-2}$. The $D_{25}$ ellipse is also
shown. North is to the top and East is to the left.}
\label{sources}
\end{figure*}

\begin{table}
\caption{Details of NGC\,5253. The distance is from Freedman \etal\
(2001), other data is from the LEDA and NED databases. The infrared
luminosity $L_{IR}$ is calculated using the 12, 25, 60 and $100\mu$m
IRAS fluxes (Sanders \& Mirabel 1996) and the H$\alpha$ luminosity is
from Marlowe \etal\ (1997, corrected for distance). The corresponding
star-formation rates (SFR) for both the IR and H$\alpha$ luminosities
are calculated using the conversion formulae in Kennicutt (1998).}
\begin{center}
\begin{tabular}{ll} \hline
Parameter & Value \\ \hline
Classification\ \ \ \ \  & Im pec, H{\small II} \\
Distance & 3.15Mpc \\
RA (J2000) & 12 39 56.0 \\
Dec (J2000) &$-31$ 38 36\\
$D_{25}$ ellipse & $4.8' \times 1.9'$\\
Position angle & $45^\circ$\\
$m_B$ & 10.75\\
$B-V$ & 0.43\\
$L_{IR}$ & $4.5\times 10^{42}\ergs$\\
$L_{\rm {H}\alpha}$ & $2.6\times 10^{40}\ergs$\\
SFR rate (IR) & $0.2\msunyr$\\
SFR rate (H$\alpha$) & $0.2\msunyr$\\ \hline
\end{tabular}
\end{center}
\end{table}

\begin{table}
\begin{center}
\caption{Positions and count rates of the 31 sources detected in the
NGC\,5253 {\it Chandra} S3 chip data. Column 1 is the source number ordered
in increasing R.A.. Columns 2 and 3 give the R.A. and Dec. of each source
and Column 4 lists their background subtracted count rates.}
\begin{tabular}{|c|c|c|c|} \hline
Source & RA (h m s) & Dec ($^{\circ} \; {'}\; {''}$) & Count Rate \\
 & & & ($\times 10^{-3}$ cts s$^{-1}$) \\ \hline
1 & 13 39 37.73 & -31 40 15.8 & $0.77\pm0.15$ \\
2 & 13 39 39.16 & -31 36 06.9 & $0.32\pm 0.09$ \\
3 & 13 39 39.17 & -31 36 09.2 & $0.42\pm 0.10$ \\
4 & 13 39 43.15 & -31 39 20.2 & $1.07\pm 0.16$\\
5 & 13 39 45.12 & -31 36 58.2 & $0.37\pm 0.10$\\
6 & 13 39 45.53 & -31 42 06.1 & $0.39\pm 0.10$\\
7 & 13 39 46.69 & -31 42 28.0 & $0.43\pm 0.11$\\
8 & 13 39 47.27 & -31 41 40.7 & $1.95\pm 0.22$\\
9 & 13 39 50.66 & -31 39 20.7 & $1.50\pm 0.18$\\
10 & 13 39 51.17 & -31 38 40.8 & $0.32\pm 0.09$\\
11 & 13 39 51.62 & -31 40 16.2 & $0.48\pm 0.11$\\
12 & 13 39 51.96 & -31 41 43.9 & $4.03\pm 0.30$\\
13 & 13 39 52.07 & -31 38 59.7 & $0.19\pm 0.07$\\
14 & 13 39 53.36 & -31 43 43.9 & $1.42\pm 0.20$\\
15 & 13 39 53.80 & -31 38 32.7 & $0.66\pm 0.13$\\
16 & 13 39 55.39 & -31 38 32.8 & $0.80\pm 0.17$\\
17 & 13 39 55.69 & -31 38 31.5 & $1.02\pm 0.20$\\
18 & 13 39 55.85 & -31 38 22.8 & $1.03\pm 0.20$\\
19 & 13 39 56.04 & -31 38 24.8 & $1.43\pm 0.23$\\
20 & 13 39 56.15 & -31 38 35.9 & $0.58\pm 0.16$\\
21 & 13 39 56.35 & -31 38 20.5 & $8.22\pm 0.43$\\
22 & 13 39 56.45 & -31 38 25.8 & $1.53\pm 0.21$\\
23 & 13 39 56.67 & -31 42 45.3 & $1.57\pm 0.21$\\
24 & 13 39 57.43 & -31 37 32.5 & $1.50\pm 0.06$\\
25 & 13 39 59.27 & -31 37 50.2 & $0.33\pm 0.09$\\
26 & 13 40 02.19 & -31 37 19.1 & $0.96\pm 0.15$\\
27 & 13 40 03.40 & -31 37 14.5 & $0.78\pm 0.13$\\
28 & 13 40 03.84 & -31 44 03.7 & $1.25\pm 0.21$\\
29 & 13 40 04.29 & -31 37 30.3 & $0.41\pm 0.10$\\
30 & 13 40 04.97 & -31 42 03.0 & $0.24\pm 0.08$\\
31 & 13 40 10.81 & -31 40 33.2 & $0.29\pm0.09$\\ \hline
\end{tabular}
\end{center}
\end{table}

\begin{figure*}
\vspace{13cm} 
\includegraphics{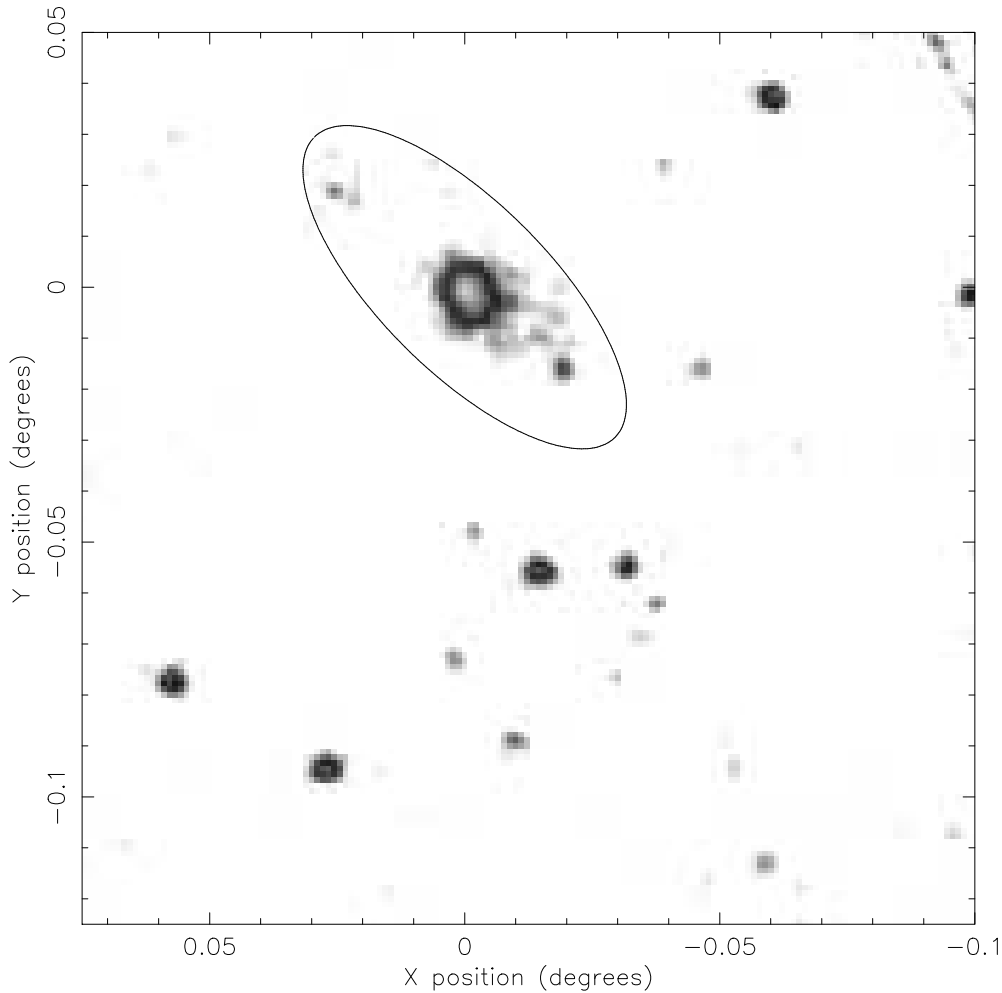}
\caption{Low resolution (smoothed using a Gaussian with FWHM of 2 pixels 
$\sim 8^{''}$), {\it XMM-Newton}, background subtracted and exposure
corrected, mosaiced image of the data from the EPIC MOS1, MOS2 and PN
cameras, in the $0.3 - 8.0\kev$ energy band. The image is an inverted
log greyscale with the flux density ranging from $5.2 \times
10^{-14}\ergs$~cm$^{-2}$~arcmin$^{-2}$ to $2.1 \times
10^{-12}\ergs$~cm$^{-2}$~arcmin$^{-2}$, with North to the top and East
to the left. The $D_{25}$ ellipse is also shown. The X-ray emission is
extended towards the WSW.}
\label{mosaic}
\end{figure*}

\section{Observations and Analysis}

A 57~ks {\it Chandra} observation of NGC\,5253 was obtained on Jan
$13^{th} - 14^{th}$ 2001 and a 46~ks {\it XMM-Newton} observation was
obtained on $8^{th} - 9^{th}$ August of the same year. For the {\it
Chandra} data, analysis was carried out on the data contained within the
S3 chip of the ACIS$-$S instrument using {\it CIAO} (version 2.2.1),
{\it HEASOFT} (version 5.1), {\it XSPEC} (version 11.1.0) and {\it
ASTERIX} (version 2.3-b1). The data were first reprocessed using CALDB
version 2.9 and then filtered to remove periods of flaring and lower
than average count rates using the lc$\_$clean.sl script 
with a $5\sigma$ clipping, leaving a total of
47.8~ks of useful data.  The data were then further filtered to retain
only data in the energy band from $0.3 - 8.0\kev$ and were background
subtracted using the appropriately scaled {\it Chandra} X-ray Observatory
Center (CXC) background event data
set. After production of an exposure map, the point sources
present in the data were detected using the  {\it CIAO wavdetect} tool, 
run with an exposure map and a
source significance threshold of $9.5 \times 10^{-7}$, a value of
$\sim$~(number of pixels)$^{-1}$ which should have limited the number of
false detections to $\sim 1$. A total of 31 point sources were detected
and these are shown in Fig.~\ref{sources}, overlaid on a smoothed image
of the ACIS-S3 chip (smoothed with a Gaussian having a FWHM of 4 pixels
$\sim2{''}$), while their positions and background subtracted count
rates are listed in Table~2. Of these 31 sources, 17 lie
within the optical extent of the galaxy, as shown by the $D_{25}$
ellipse (de Vaucouleurs \etal\ 1991) on Fig.~\ref{sources}. These 17
sources are the ones most likely to be associated with NGC\,5253 and so
are the only ones considered further in this analysis. We note that
results from this data have already been presented as part of a survey
of the point source populations of nearby galaxies by Colbert \etal\
(2004), and a comparison with these results will be presented later.

\begin{figure}
\vspace{7.0cm} 
\includegraphics{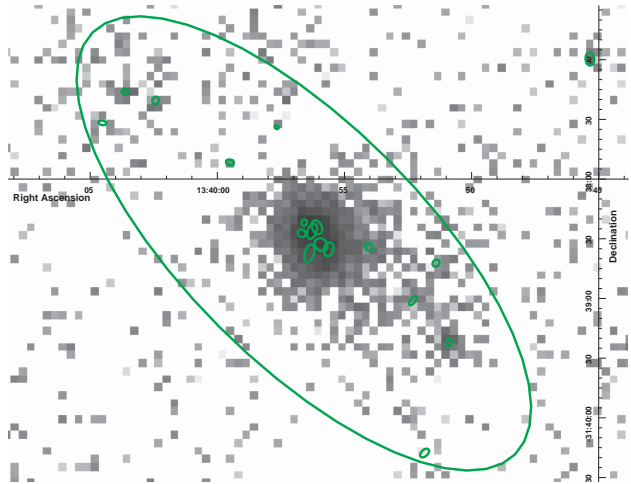}
\caption{Unsmoothed background subtracted and exposure
corrected, mosaiced image of the data from the {\it XMM-Newton} EPIC MOS1,
MOS2 and PN cameras in the $0.3 - 8.0\kev$ energy range. The overlaid small
ellipses show the positions and sizes of the sources detected in the {\it
Chandra} data. The $D_{25}$ ellipse is also shown. 
The extension in the X-ray emission towards the WSW is
largely associated with point sources, but the emission is also slightly
extended to the NW and SE along the galaxy's minor axis, as well as along
the major axis. North is to the top and East is to the left.}
\label{xmmchan}
\end{figure}

From the {\it XMM-Newton} observation, only the European Photon Imaging
Camera (EPIC) data has been
analysed. Removal of flares using a recursive $3\sigma$ clipping technique
and standard pattern and flag filtering for the three cameras left 39.0~ks
of useful data for MOS1, 39.6~ks for MOS2 and 26.7~ks for PN. These data
were then also filtered to retain only data in the $0.3 - 8.0\kev$ energy
band in line with the {\it Chandra} data. A background subtracted (using
the David Lumb `blank sky' and Philippe Marty `closed' events files) mosaic
of the exposure corrected images from the three cameras is shown in
Fig.~\ref{mosaic}. The (0,0) point on this image lies at the centre of the
$D_{25}$ ellipse at a position of $\alpha = 13^{h}39^{m}56.2^{s}$, $\delta
= -31^{\circ}38{'}29.9{''}$. The diffuse X-ray emission is seen to be
extended to the WSW of the nucleus which is not apparent in the {\it
Chandra} image. The latter does however show evidence of absorption of the
X-ray emission due to the presence of the dust lane that extends E from the
nuclear region, which is not apparent on the {\it XMM-Newton} data. It is
also apparent from comparing Figs.~\ref{sources} and \ref{mosaic} that some
of the sources within the {\it Chandra} ACIS-S3 field-of-view are
exhibiting variability. In the eight months between the two observations,
sources 25 and 30 have faded whilst six possible additional sources are
apparent in the {\it XMM-Newton} data, to the east of source 1, south-east
of source 7, north-east of source 12, north of source 24, north of source
26 and north of source 27. The last three in this list may be associated
with NGC\,5253 as they lie just inside the $D_{25}$
ellipse. Fig.~\ref{xmmchan} shows the position and extent of the {\it
Chandra} sources overlaid on the unsmoothed {\it XMM-Newton} mosaic and
shows that a large fraction of what appears to be extended diffuse emission
to the SW along the galaxy's major axis is more likely to be associated
with point sources. However the diffuse emission does show extension to
both the NW and SE along the minor axis, which may be associated with the
shells and radial filaments seen in H$\alpha$ images (Marlowe \etal\ 1995).

After detection, the point sources were subtracted from the {\it Chandra}
data so that the diffuse X-ray emission contained within that observation
could be investigated. Also, a 3 colour adaptively smoothed image of the
central region of the galaxy has been produced using {\it csmooth} in fft
(fast Fourier transform) mode and this is shown in Fig.~\ref{3col}. The
lower and upper sigmas for deriving the smoothing kernel were set at 2 and
5 respectively.  This image shows that the spectral characteristics of the
sources vary and that the diffuse emission appears to consist mainly of
soft emission lying behind regions if varying absorption. The red, green
and blue images used correspond to energy bands of $0.3-1.1\kev$,
$1.1-3.8\kev$ and $3.8-8.0\kev$ respectively. These bands were chosen so
that the total number of counts in each band was approximately equal and
thereby maximising the S/N ratio for the bands. The diffuse emission is
seen to be confined predominantly to the nuclear region of the galaxy and
the position of the indentation seen to the East, suggesting heavy
absorption, corresponds to the position of the dust lane identified in the
galaxy (Hunter 1982). In addition the effect of the dust is seen, in the
reduction of intensity and change of colour as a result of the absorption
of the softer emission, in the darker band running from E to W across the
nuclear region.

\section{Results}

\subsection{Integrated Emission}

The background subtracted spectrum of the total X-ray emission from
within the $D_{25}$ ellipse of NGC\,5253 was fitted separately for the
four instruments (ACIS-S, MOS1, MOS2 and PN), using the modified
Levenberg-Marquardt method and standard $\chi^{2}$ statistic from {\it
XSPEC}, with an absorbed 2 thermal component plus power-law fit, the
same model that was fitted in earlier work to the similar dwarf
starburst galaxy NGC\,4449 (Summers \etal\ 2003). It was found though,
for NGC\,5253 that each component had to be fitted with a separate
absorbing column, in addition to the absorption due to the interstellar
medium within the Milky Way (Galactic column density, $N_{H(GAL)} = 3.87
\times 10^{20}$~cm$^{-2}$), to obtain a satisfactory fit. Another
difference in the fitting for these two galaxies was in the energy range
over which the spectral fitting was carried out. The NGC\,5253 {\it
Chandra} data was found to contain both a hard excess above $\sim
6.0\kev$ and a soft excess below $\sim 0.37\kev$, neither of which could
be adequately dealt with by either the flare removal or background
subtraction. Even the use of a local background instead of the
background event data set did not remove them and so the spectral
fitting for all the instruments was confined to the $0.37 -6.0\kev$
energy band for ease of comparison of the results from the two different
telescopes. In addition, a single simultaneous fit to the data from the
three {\it XMM-Newton} cameras was performed. The two absorbed thermal
components were modelled using the {\it wabs} (photo-electric absorption
using Wisconsin cross-sections) and {\it mekal} thermal plasma codes
within {\it XSPEC}. Initially, Galactic absorption was assumed for all
the column densities and a value of $0.19~Z_{\odot}$ assumed for the
metallicity of the X-ray emitting gas (Ferrarese \etal\ 2000). The
results are summarised in Table~3 and the fitted spectra for the {\it
Chandra} data and simultaneous fit to the {\it XMM-Newton} data are
shown in Fig.~\ref{totspec}. Two thermal plasmas were used to represent
the diffuse emission firstly because multi-phase models are seen to be
needed to best fit the emission from other starburst galaxies
(e.g. NGC\,253, Strickland \etal\ 2002; NGC\,1569, Martin \etal\ 2002;
NGC\,4449, Summers \etal\ 2003), and secondly because single temperature
fits to just the diffuse emission do not have a wide enough energy
distribution to fit the spread seen in the data and so are statistically
less robust.

\begin{figure*}
\vspace{11cm}  
\includegraphics{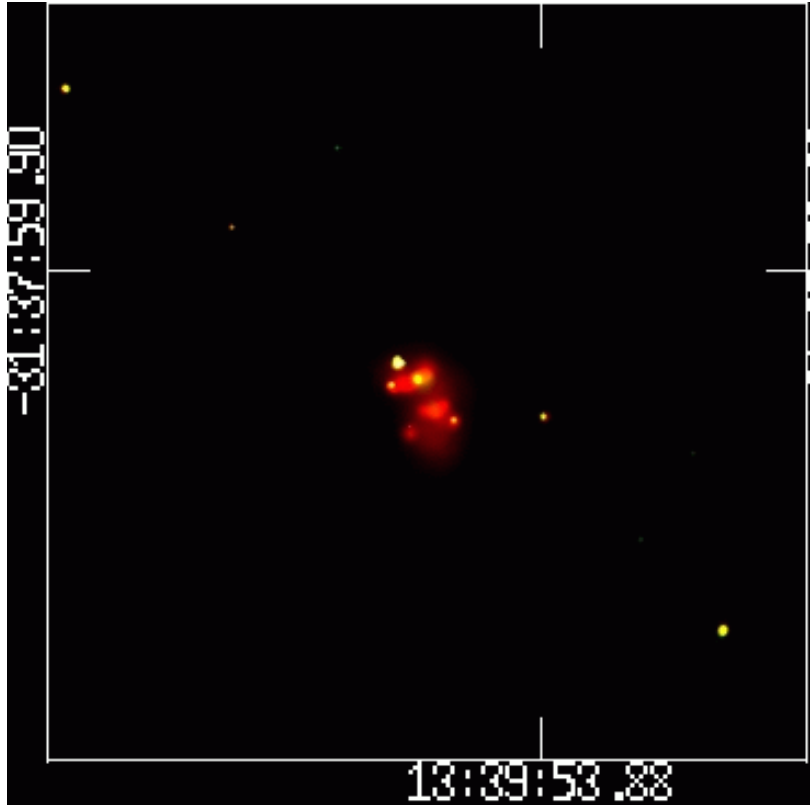}
\caption{Adaptively smoothed 3 colour image of a 3 arcmin square region of
the {\it Chandra} ACIS-S3 chip, centred on the nucleus of NGC\,5253 at
$\alpha = 13^{h}39^{m}56.2^{s}$, $\delta = -31^{\circ}38{'}29.9{''}$ (red:
$0.3-1.10\kev$, green: $1.10-3.80\kev$ and blue: $3.80-8.0\kev$). The
hardness variation in the source spectra is evident, while the 
diffuse emission is seen to be predominantly soft.}
\label{3col}
\end{figure*}

\begin{table*}
\begin{center}
\caption{Best fits to the background subtracted spectra of the total X-ray
emission from within the $D_{25}$ ellipse of NGC\,5253 for a
$wabs_{GAL}(wabs(mekal)+wabs(mekal)+wabs(po))$ model, with the Galactic
column $wabs_{GAL}=3.87\times 10^{20}$~cm$^{-2}$. The 
five columns of
results proceeding from left to right are for the {\it Chandra} ACIS-S3
chip, the individual {\it XMM-Newton} MOS1, MOS2 and PN cameras and the
combined results from the data for the three {\it XMM-Newton} cameras fitted
simultaneously. Row 1 lists the individual instruments, rows 2, 4, and 7
are the fitted column densities for the three components, $N_{H(1)}$,
$N_{H(2)}$ and $N_{H(3)}$ respectively. Rows 3 and 5 are the temperatures,
$kT_{1}$ and $kT_{2}$ of the two thermal components and row 6 is the fitted
metallicity of the X-ray emitting gas, fitted simultaneously for the two
thermal components. Row 8 gives the photon index, $\Gamma$, of the 
power-law component, while row 9 gives the fit statistics.}
\begin{tabular}{|l|c|c|c|c|c|} \hline
           & {\it Chandra} & \multicolumn{4}{c}{\it XMM-Newton}  \\
  \cline{3-6} 
Instrument & ACIS-S & MOS1 & MOS2 & PN & 3 XMM \\ \hline
$N_{H(1)}$ ($\times 10^{21}$~cm$^{-2}$)  & $1.01\pm^{0.12}_{0.08}$ &
$1.00\pm^{0.43}_{0.32}$ & $0.89\pm^{0.31}_{0.20}$ &
$0.79\pm^{0.12}_{0.11}$ & $0.48\pm^{0.09}_{0.07}$ \\  
$kT_{1}$ (\kev) & $0.25\pm 0.01$ & $0.26\pm0.02$ &
$0.22\pm^{0.02}_{0.01}$ & $0.22\pm 0.01$ & $0.25\pm 0.01$ \\ 
$N_{H(2)}$  ($\times 10^{21}$~cm$^{-2}$) & $3.84\pm^{0.34}_{0.23}$ &
$2.06\pm^{0.67}_{0.51}$ & $1.34\pm^{0.70}_{0.32}$ &
$0.82\pm^{0.28}_{0.24}$ & $3.78\pm^{0.36}_{0.35}$ \\ 
$kT_{2}$ (\kev) & $0.76\pm 0.04$ & $0.64\pm0.08$ &
$0.59\pm^{0.06}_{0.09}$ & $0.73\pm^{0.06}_{0.06}$ &
$0.67\pm^{0.04}_{0.05}$ \\ 
$Z$  ($Z_{\odot}$) & $0.13\pm0.01$ & $0.16\pm0.02$ & $0.11\pm0.01$ &
$0.13\pm0.01$ & $0.11\pm0.01$ \\ 
$N_{H(3)}$ ($\times 10^{21}$~cm$^{-2}$) &  $46.4\pm^{6.04}_{4.59}$ &
$0.51\pm^{1.44}_{0.51}$ & $5.00\pm^{13.10}_{3.64}$ &
$8.77\pm^{7.31}_{4.11}$ & $7.01\pm^{30.09}_{5.34}$ \\  
$\Gamma$ & $1.69\pm^{0.06}_{0.06}$ & $3.60\pm^{0.78}_{1.19}$ &
$2.12\pm^{1.68}_{0.62}$ & $1.92\pm^{0.47}_{0.33}$ &
$1.54\pm^{0.82}_{0.39}$ \\ 
$\chi^{2}$/d.o.f. & $206/150$ & $66/53$ & $71/66$ & $159/145$ &
$323/284$ \\ \hline 
\end{tabular}
\end{center}
\end{table*}

\begin{figure*}
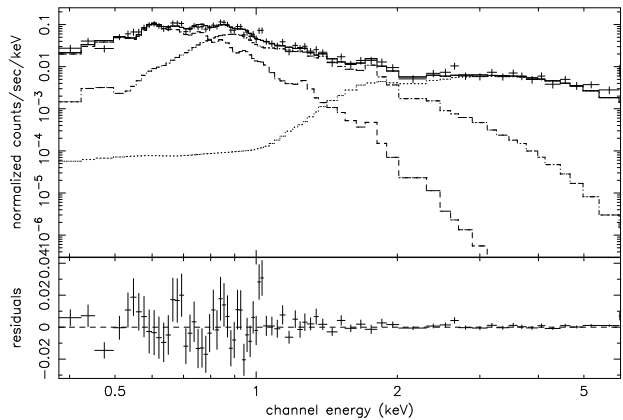

\vspace{6.0cm}
\includegraphics{n5253_fig5a.ps}
\includegraphics{n5253_fig5b.ps}
\caption{Fitted spectrum of the total X-ray emission from NGC~5253 Left
panel: {\it Chandra} data, right panel: simultaneous fit to the data from
the three {\it XMM-Newton} cameras. The fit is for an absorbed two
temperature plus power-law fit, with a column density of 
$N_{H(GAL)}=3.87 \times
10^{20}$~cm$^{-2}$ assumed for Galactic absorption. See
Table~3 and text for details of the individual fits. The individual
contributions from the separate components are shown on the {\it
Chandra} data.} 
\label{totspec}
\end{figure*}

We note that the MOS1 data gives substantially different results from
the other two {\it XMM-Newton} cameras, particularly for the column
densities and metallicities. There is also a difference between the
results obtained from the two satellites, with the {\it Chandra} data
giving higher column densities for the soft thermal and for the
power-law components. The average unabsorbed flux, in the $0.37-6.0\kev$
energy band, from NGC\,5253 for the {\it Chandra} data is
$(1.56\pm^{0.11}_{0.12})\times 10^{-12}\ergs$~cm$^{-2}$, and for the
combined {\it XMM-Newton} data is $(3.22 \pm^{0.27}_{0.40}) \times
10^{-13}\ergs$~cm$^{-2}$. These values correspond to a total intrinsic
X-ray luminosity of $(1.86 \pm^{0.13}_{0.14})\times 10^{39}\ergs$ from
the {\it Chandra} data and $(3.83 \pm^{0.33}_{0.47}) \times
10^{38}\ergs$ from the {\it XMM-Newton} data for our assumed distance of
3.15~Mpc. In both cases, these values are higher than the results
obtained from earlier {\it ROSAT} and {\it Einstein} data but the
differences are explained by the higher fitted column densities of the
more recent data. The fraction of the emission due to the individual
components is $15\%$ - cooler thermal, $25\%$ - hotter thermal and
$60\%$ - power-law from the {\it Chandra} data and $40\%$ - cooler
thermal, $50\%$ - hotter thermal and $10\%$ - power-law from the {\it
XMM-Newton} data. The very high fitted column density for the power-law
component in the {\it Chandra} data accounts for a lot of this
discrepancy. The differences in the calibration between the MOS1 camera
and the other two {\it XMM-Newton} cameras will also lead to possible
discrepancies when comparing results from the combined {\it XMM-Newton}
fits to {\it Chandra} fits. We do however note the {\it Chandra}
calibration problems at low energy, which may affect matters at low
energies. An important point is that that the discrepancies between fits
to the {\it XMM-Newton} and {\it Chandra} data are much smaller for the
two thermal components, and it is these components that we concentrate on
in this paper.

\subsection{Point Sources}

\begin{figure}
\vspace{6.0cm}
\includegraphics{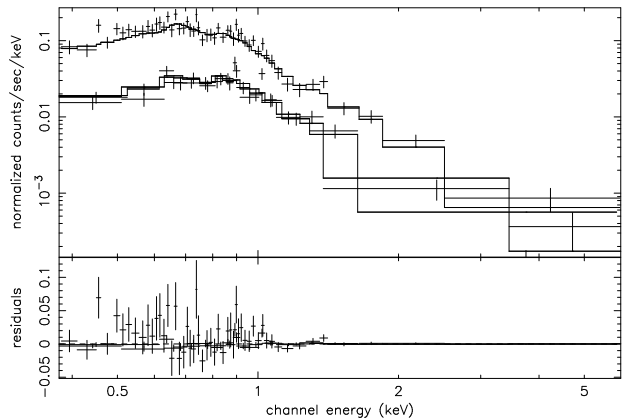}
\caption{Spectral fit obtained for source 21, the most luminous source
in the {\it Chandra} data. The fit is an absorbed two component
(mekal+power-law) fit with the Galactic column density assumed to be $3.87
\times 10^{20}$~cm$^{-2}$ and the metallicity fixed at $0.13 Z_{\odot}$
(the average fitted value from the fits to the total emission). The best
fit gives $N_{H}=(8.56 \pm^{0.40}_{0.34}) \times 10^{21}$~cm$^{-2}$,
$kT= 0.24 \pm 0.01\kev$ and the photon index, $\Gamma =
2.09\pm^{0.19}_{0.15}$.}
\label{src21}
\end{figure}

A background subtracted spectrum was extracted from the {\it Chandra}
data for each of the 17 sources within the $D_{25}$ ellipse. (The
background spectra were taken from the aforementioned CXC background
file and were appropriately scaled. The use of the background files
meant that the background spectrum was taken from the identical position
on the CCD and avoided the problem of contamination from adjacent
sources, which as shown in Fig.~\ref{sources} is particularly a problem
in the central region of NGC\,5253.) The spectra were grouped so that
each had a minimum of 5 data bins and when possible a minimum of 10
counts per bin. (In fact, due to the low statistics, this was only
possible in the case of sources 9, 19, 21 and 22.) No attempt was made
initially to fit the data where a source had less than 50 counts after
background subtraction. Source 21 has the highest count-rate by far of
any of the sources (see Table~2) and is fitted well by an
absorbed two component (mekal plus power-law) fit. Galactic absorption
is assumed to be $N_{H(GAL)}=3.87 \times 10^{20}$~cm$^{-2}$ and the
metallicity of the X-ray emitting gas is fixed at $0.13Z_{\odot}$ 
[the average value fitted to the total X-ray emission from
NGC\,5253, which is slightly lower than, but agrees within errors with,
the value of $0.19\pm^{0.08}_{0.06} Z_{\odot}$ obtained for the H{\small
II} regions in the centre of NGC\,5253 by Webster \& Smith (1983)]. The
fitted spectrum for this source is shown in Fig.~\ref{src21} and the
fitted parameters are summarised in Table~4 along with those
for the other 16 sources within the $D_{25}$ ellipse. Where possible,
the other sources were fitted with this same model unless the
normalisation went to zero for one of the components, in which case that
component was omitted and these fits are indicated by a dash in the
parameter values in Table~4. As suggested by Strickland \&
Stevens (1999), from analysis of the {\it ROSAT} HRI data for this
galaxy, some of the sources detected in the central region of NGC\,5253
may be due to superbubbles surrounding super star-clusters rather than
individual point sources, such as X-ray binaries. Two of these five
sources found by Strickland \& Stevens (1999) have been further resolved
into two sources using the superior resolution of {\it Chandra}. We can
relate the HRI sources recorded in Strickland \& Stevens (1999) to the
{\sl Chandra} sources found here as follows: {\sl ROSAT} source A
corresponds to source 21, B to source 22, C to source 20, D to sources
18/19 and E to sources 16/17.  The assertion that these sources could be
extended superbubbles is supported by the source detections here where
in particular, sources 16, 17, 18, 19 and 20 are all detected as being
elliptical whereas the PSFs at the corresponding positions on the CCD
are all circular. The inclusion of a thermal component in the spectral
fits for these sources is then to be expected and in the cases of
sources 16, 17, 18 and 20, the thermal components completely dominate
the spectra. Source 21 appears to be substantially more luminous in this
{\sl Chandra} observation than in the {\sl ROSAT} HRI observation,
though a detailed comparison is difficult because of the low source
counts in the HRI observation.

\begin{figure}
\vspace{8.0cm}  
\includegraphics{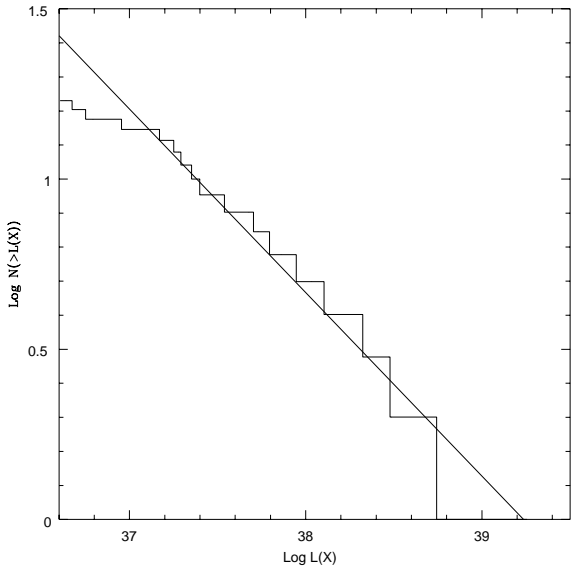} 
\caption{$\log(N)-\log(L_{X})$ plot for the 17 point sources detected
within the $D_{25}$ ellipse of NGC\,5253. The $L_{X}$ values are
absorption corrected and have been fitted as shown in Table~4.
The line shown is the power-law fit to the high luminosity end of the
data, with a slope of $-0.54 \pm^{0.21}_{0.16}$. Incompleteness occurs
at an absorption corrected luminosity of $1.46 \times 10^{37}\ergs$.}
\label{logn}
\end{figure}

From the best fits to all the sources within the $D_{25}$ ellipse, the
absorption corrected fluxes and luminosities in the $0.3 - 8.0\kev$
energy band were determined. These luminosities are also included in
Table~4 and the $\log(N)-\log(L_X$) plot for the 17 sources possibly
associated with NGC\,5253 is shown in Fig.~\ref{logn}. This has been
fitted with a single power-law, with a slope of $-0.54
\pm^{0.21}_{0.16}$ for the higher luminosity sources. Sources with
luminosities below an absorption corrected luminosity of $1.46 \times
10^{37}\ergs$ were not included in the fit, as these figures represent
the completeness limit for the data. The sources where the thermal
component dominates are included in this data, even though these may
represent the emission from young superbubbles rather than point
sources. The slope determined for all the sources is not quite as steep
(though comparable to within the errors) as the $-0.70$ found for
NGC\,4214 (Hartwell \etal\ 2003) but slightly steeper than the $-0.51$
found for NGC\,4449 (Summers \etal\ 2003). As a further comparison,
Martin \etal\ (2002) give an upper limit of $-0.55$ for the slope of the
luminosity function of sources detected in the $1.1-6.0\kev$ energy band
in NGC\,1569, but also state that the value is not well constrained. The
low source counts in the NGC\,5253 data make anything more rigorous than
a statement of the slope values impossible. The NGC\,5253 value is
steeper than the value quoted ($\sim -0.45$) by Zezas \etal\ (2001) for
the larger starburst galaxies, M82 and the Antennae.  The steeper slopes
seen in the dwarfs may be the result of them having fewer high
luminosity sources, and in the case of NGC\,5253 this could be directly
related to it being a younger starburst. (For more analysis of the
luminosity functions of these and other starburst galaxies see Hartwell
\etal\ 2003). The effect of omitting the sources dominated by thermal
emission which are probable young superbubbles (sources 16, 17, 18, 19
and 20) is to decrease the slope of the fit to $-0.41$, which is more in
line with the other results quoted above. This highlights the fact that
avoiding source confusion is crucial to this type of analysis and the
inclusion of super star-clusters within superbubbles, as point sources
biases the results in such a way as to steepen the slope of the point
source luminosity function.

We note that Colbert \etal\ (2004) in their analysis of the NGC\,5253
data, quote a steeper value for the point source luminosity function of
$-0.92$. A major difference between this and that analysis is that here
we fit the spectrum of each source individually (Table~4), whereas
Colbert \etal\ (2004) determine source fluxes using a power-law model
with a fixed slope and a column set at the Galactic absorption
level. This may well explain the different slopes determined for
NGC\,5253.

The data were point source searched in the three different energy bands
used to produce the three colour image, (soft $0.3-1.1\kev$, medium
$1.1-3.8\kev$ and hard $3.8-8.0\kev$) and in the three energy bands used
for the analysis of NGC\,4449 (soft $0.3-0.8\kev$, medium $0.8-2.0\kev$
and hard $2.0-8.0\kev$) so that a direct comparison could be made of the
sources in the two galaxies. The hardness ratios for the individual
sources within the $D_{25}$ ellipse were calculated for the sources
detected in more than one of the energy bands. Due to the low count
rates for many of the sources most were detected in only 1 or 2
bands. The total counts in each of these energy bands and the calculated
hardness ratios are shown in Tables~5 and 6, for
the two different sets of energy bands. Note, that when point source
searching was carried out in the separate energy bands, sources 18 and
19 when detected as a single source rather than two sources as in the
broadband searching. The two hardness ratios shown are $(m - s)/(m + s)$
and $(h -m)/(h +m)$, where $s$, $m$ and $h$ are the counts in the soft,
medium and hard bands respectively. It is noticeable, particularly in
Table~5, that the sources in the central region (sources
16-20) are all very soft, with very few counts above 1.0\kev and
virtually none above 2.0\kev, suggesting that these are associated with
the emission from superbubbles surrounding super star-clusters rather
than individual harder objects like X-ray binaries. 

\begin{table*}
\begin{center}
\caption{Results of fitting a {\it wabs$_{GAL}$(wabs(mekal+po))} model (the 
best fit model for source 21) to all 17 sources within the $D_{25}$
ellipse. Column 1 gives the source numbers as shown on
Fig.~\ref{sources}. Column 2 contains the fitted column density for each
source assuming a Galactic column density of $N_{H(GAL)}=3.87 \times
10^{20}$~cm$^{-2}$. Columns 3 and 4 contain the temperature of the {\it
mekal} component and photon index of the power-law component
respectively. Column 5 and 6 show the absorption corrected flux and
luminosity for each source. The errors shown are the 90\% confidence regions
for each parameter.}
\begin{tabular}{|c|c|c|c|c|c|} \hline
Source & $N_{H}$ & $kT$ & $\Gamma$ & Unabsorbed Flux & Unabsorbed Luminosity\\
 & ($\times 10^{21}$~cm$^{-2}$) & (\kev) &  & ($\times
10^{-14}\ergs$~cm$^{-2}$) & ($\times 10^{38}\ergs$) \\ \hline
9 &  $4.48\pm^{1.24}_{0.91}$ & $0.08\pm^{0.01}_{0.01}$ &
     $1.94\pm^{0.40}_{0.28}$ & $25.7\pm^{15.0}_{14.8}$ &
     $3.06\pm^{1.79}_{1.76}$ \\
10 & $7.91\pm^{76.04}_{7.91}$ & $0.25\pm^{0.07}_{0.17}$
   & $1.99\pm^{0.73}_{0.36}$  & $0.52\pm^{0.61}_{0.34}$ &
     $0.06\pm^{0.07}_{0.04}$  \\
11 & $7.84\pm^{4.31}_{1.82}$ & $0.23\pm^{0.06}_{0.09}$ &
     $1.45\pm^{0.84}_{0.41}$ & $1.70\pm^{1.40}_{1.40}$ &
     $0.20\pm^{0.17}_{0.17}$ \\
13 & $44.59\pm^{24.53}_{7.73}$ & $0.25\pm^{0.03}_{0.08}$ &
     $1.52\pm^{4.25}_{0.47}$ & $6.10\pm^{5.50}_{1.60}$ &
     $0.73\pm^{0.66}_{0.19}$ \\
15 & $5.02\pm^{1.84}_{1.10}$ & $0.25\pm^{0.07}_{0.07}$ &
     $2.99\pm^{1.37}_{0.60}$ & $1.40\pm^{0.90}_{0.90}$ &
     $0.17\pm^{0.11}_{0.11}$ \\
16 & $6.18\pm^{1.04}_{0.76}$ & $0.17\pm^{0.01}_{0.01}$ &
-- & $12.6\pm^{4.10}_{3.60}$ & $1.50\pm^{0.49}_{0.43}$ \\
17 & $5.52\pm^{1.21}_{0.79}$ & $0.25\pm^{0.02}_{0.02}$ &
-- & $4.00\pm^{1.10}_{1.10}$ & $0.48\pm^{0.13}_{0.13}$ \\
18 & $5.33\pm^{1.00}_{0.64}$ & $0.24\pm^{0.02}_{0.02}$ &
-- & $4.50\pm^{1.20}_{1.20}$ & $0.54\pm^{0.14}_{0.14}$ \\
19 & $6.84\pm^{0.99}_{0.75}$ & $0.24\pm^{0.02}_{0.02}$ &
     $1.41\pm^{0.93}_{0.44}$ & $9.10\pm^{2.40}_{2.40}$ &
     $1.08\pm^{0.29}_{0.29}$ \\
20 & $3.99\pm^{1.16}_{0.75}$ & $0.22\pm^{0.02}_{0.02}$ & 
-- & $2.10\pm^{0.70}_{0.70}$ & $0.25\pm^{0.08}_{0.08}$ \\
21 & $8.56\pm^{0.40}_{0.34}$ & $0.24\pm^{0.01}_{0.01}$ &
     $2.09\pm^{0.19}_{0.15}$ & $85.3\pm^{10.5}_{10.8}$ &
     $10.0\pm^{1.30}_{1.30}$ \\
22 & $5.35\pm^{0.70}_{0.50}$ & $0.15\pm^{0.01}_{0.01}$ &
     $2.50\pm^{5.23}_{1.68}$ & $24.9\pm^{6.40}_{6.50}$ &
     $2.96\pm^{0.76}_{0.77}$ \\
24 & $1.43\pm^{6.42}_{1.04}$ & $0.08\pm^{0.02}_{0.02}$ &
     $1.22\pm^{2.83}_{0.58}$ & $0.37\pm^{0.44}_{0.35}$ &
     $0.04\pm^{0.05}_{0.04}$ \\
25 & $2.28\pm^{3.88}_{1.78}$ & $0.18\pm^{0.06}_{0.10}$ &
     $2.50\pm^{3.80}_{1.21}$ & $0.43\pm^{0.55}_{0.35}$ &
     $0.05\pm^{0.07}_{0.04}$ \\
26 & $0.84\pm^{1.33}_{0.68}$ & $0.15\pm^{0.02}_{0.05}$ &
     $0.86\pm^{1.39}_{0.79}$ & $2.10\pm^{0.80}_{1.00}$ &
     $0.25\pm^{0.10}_{0.12}$ \\
27 & $4.56\pm^{10.75}_{4.56}$ & -- &
     $1.12\pm^{0.61}_{0.58}$ & $1.10\pm^{0.40}_{0.40}$ &
     $0.13\pm^{0.05}_{0.05}$ \\
29 & $7.13\pm^{3.39}_{1.80}$ & $0.29\pm^{0.07}_{0.07}$ & 
     $0.70\pm^{3.28}_{0.82}$ & $1.60\pm^{1.10}_{1.10}$ &
     $0.19\pm^{0.13}_{0.13}$ \\ \hline
\end{tabular}
\end{center}
\end{table*}

\begin{figure}
\vspace{7cm}  
\includegraphics{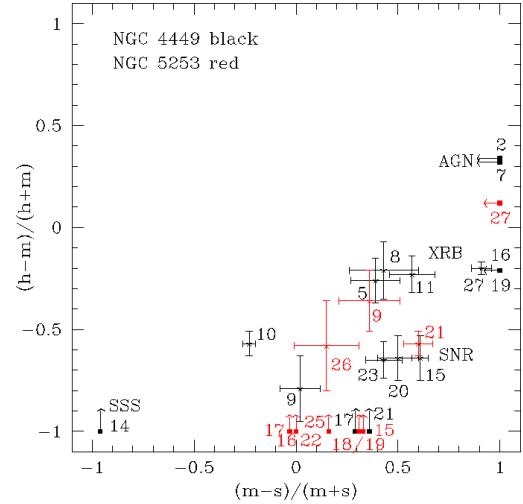} 
\caption{Hard vs. soft hardness ratios for the sources detected in the
separate energy bands. The data shown in black are for NGC\,4449 
(Summers \etal\ 2003) and in red for NGC\,5253. The hard hardness
ratio is calculated from the $2.0-8.0\kev$ and $0.8-2.0\kev$ energy bands
and the soft hardness ratio is calculated from the $0.8-2.0\kev$ and
$0.3-0.8\kev$ energy bands, in the usual way, as defined in the headings of
columns 5 and 6 of Table~6. An indication of the likely
origin of the 
sources depending on their location in this plot is shown by labels placed
next to the identified sources in the NGC\,4449 data. The sources plotted
with squares to the right and bottom were only detected in 2
out of the 3 energy bands. The arrows indicate the
direction that the sources would move if higher count rates permitted
detections in all three energy bands. The error bars have also been omitted
from these sources to avoid additional confusion on the plot. See
Table~6 for those associated with NGC\,5253.}
\label{hr}
\end{figure}

In general, the sources detected in NGC\,5253 are softer than those in
NGC\,4449 and lie behind an on average higher absorbing column
density. The implication of this being that the sources in NGC\,5253
must be softer as the higher column density would preferentially absorb
the softer X-ray emission. The low counts for the sources in NGC\,5253
will effect these results too, but still, only 5 out of 17 sources
have detections in the $2.0 - 8.0\kev$ band compared with 15 out of 24
for NGC\,4449. To better compare some of these sources, the sources
where hardness ratios can be calculated from the NGC\,5253 data, in the
$0.3 - 0.8\kev$, $0.8 - 2.0\kev$ and $2.0 - 8.0\kev$ energy bands, have
been overlaid on the hard vs. soft hardness ratio plot for
NGC\,4449. The results are shown in Fig.~\ref{hr} where the NGC\,4449
data is shown in black and that for NGC\,5253 is in red. The sources
represented by squares with arrows, to the right and bottom of the plot,
were only detected in 2 of the 3 energy bands and so their positions
represent the most extreme positions possible for them on this
plot.

The average fitted column density for the NGC\,4449 sources is $\sim 3.0
\times 10^{21}$~cm$^{-2}$ while that for the NGC\,5253 sources is $\sim
7.5 \times 10^{21}$~cm$^{-2}$. The effect of these differing values on a
plot such as Fig.~\ref{hr} is to displace the sources towards the right
and slightly towards the top of the plot, as the absorbing column behind
which they lie increases. For a truer comparison, the NGC\,5253 plotted
values would decrease by $\sim 0.2 - 0.3$ in the $(m - s)/(m + s)$ ratio
and $\sim 0.1$ in the $(h - m)/(h + m)$ ratio, if the average column
density of NGC\,4449 was assumed. Of the 5 sources detected in the $2.0
- 8.0\kev$ band, two of them (sources 10 and 27) along with source 13,
which in addition has a very high fitted column density, have spectra
that show no emission below 1.0\kev. As such, these three sources are
possible candidates for background AGN. These figures are in good
agreement with the number of background sources that would be expected
within the $D_{25}$ ellipse from `The {\it Chandra} Deep Field South
Survey' (Campana \etal\ 2001) for our completeness limit ($\sim 2$ for
our completeness limit and up to $\sim 7$ down to a flux level
corresponding to that of our faintest detected source).

The three sources detected in all three energy bands (Sources 9, 21 and 26),
along with sources 11 and 29, which were only detected in the medium band,
are most likely X-ray binaries as none of these sources are coincident with
the positions of the reported supernova explosions in NGC\,5253 in 1895 and
1972 (1895B: $\alpha = 13^{h} 39^{m} 57.4^{s}$, $\delta = -31^{\circ} 38{'}
06{''}$; 1972E: $\alpha = 13^{h} 39^{m} 52.6^{s}$, $\delta = -31^{\circ}
40{'} 10{''}$, Caldwell \& Phillips 1989) In fact, none of the detected
sources tie up with the reported positions. Source 21 lies closest
to the reported position of SN1895B (18 arcsec to the SW of its position)
and source 11 lies closest to that for SN1972E (15 arcsec WSW of its
position).

\begin{table*}
\begin{center}
\caption{Hardness ratios for the 17 sources detected within the $D_{25}$
ellipse. Soft band, $s$, $0.3-1.1\kev$, medium band,
$m$, $1.1-3.8\kev$ and hard band, $h$, $3.8-8.0\kev$. Where no counts are
shown, {\it wavdetect} failed to detect the object in that energy
band. Column 1 gives the source numbers as shown on
Fig.~\ref{sources}. Columns 2 -- 4 are the counts in the 3 different energy
bands and columns 5 and 6 give the values of the hardness ratios calculated
as detailed in the column headings.}
\begin{tabular}{|c|c|c|c|c|c|c|c|} \hline
Source & Counts in & Counts in & Counts in & \underline{(m - s)} &
\underline{(h - m)} \\ 
 & Soft Band & Medium Band & Hard Band & (m + s) & (h + m) \\ 
& ($0.3-1.1\kev$) &  ($1.1-3.8\kev$) &  ($3.8-8.0\kev$) & & \\ \hline
9 & $31.1 \pm 5.7$ & $36.1 \pm 6.2$ & - & $0.07 \pm 0.13$ & - \\
10 & - & $6.1 \pm 2.6$ & $7.9 \pm 3.0$ & - & $0.13 \pm 0.29$ \\
11 & $7.3 \pm 2.8$ & $12.7 \pm 3.7$ & - & $0.27 \pm 0.24$ & - \\
13 & - & $7.6 \pm 2.8$ & - & - & - \\
15 & $16.0 \pm 4.2$  & $13.0 \pm 3.7$ & - & $-0.10 \pm 0.19$ & - \\
16 & $28.0 \pm 7.1$ & - & - & - & - \\ 
17 & $27.1 \pm 7.0$ & - & - & - & - \\ 
18/19 & $43.4 \pm 8.9$ & $28.7 \pm 6.5$ & - & $-0.20 \pm 0.16$ & - \\
20 & $29.7 \pm 7.6$ & - & - & - & - \\ 
21 & $182.2 \pm 14.2$ & $190.9 \pm 14.1$ & $23.0 \pm 4.9$ & $0.02 \pm 0.05$ &
$-0.79 \pm 0.09$ \\
22 & $63.3 \pm 9.2$ & $12.2 \pm 4.1$ & - & $-0.68 \pm 0.16$ & - \\
24 & - & - & - & - & - \\
25 & $7.5 \pm 2.8$ & $7.9 \pm 2.8$ & - & $0.03 \pm 0.26$ & - \\
26 & $23.0 \pm 4.9$ & $20.0 \pm 4.6$ & - & $-0.07 \pm 0.16$ & - \\
27 & - & $34.4 \pm 6.0$ & - & - & - \\
29 & $9.4 \pm 3.2$ & $9.7 \pm 3.4$ & - & $0.02 \pm 0.25$ & - \\ \hline
\end{tabular}
\end{center}
\end{table*}

\begin{table*}
\begin{center}
\caption{Hardness ratios for the 17 sources detected
within the $D_{25}$ ellipse. Soft band, $s$, $0.3-0.8\kev$, medium band,
$m$, $0.8-2.0\kev$ and hard band, $h$, $2.0-8.0\kev$. Details as per 
Table~5. These bands are used for comparison with data for NGC\,4449.}
\begin{tabular}{|c|c|c|c|c|c|c|c|} \hline
Source & Counts in & Counts in & Counts in & \underline{(m - s)} &
\underline{(h - m)} \\ 
 & Soft Band & Medium Band & Hard Band & (m + s) & (h + m) \\ 
& ($0.3-0.8\kev$) &  ($0.8-2.0\kev$) &  ($2.0-8.0\kev$) & & \\ \hline
9 & $17.3 \pm 4.2$ & $36.6 \pm 6.2$ & $17.1 \pm 4.4$ & $0.36 \pm 0.15$ &
$-0.36 \pm 0.15$ \\
10 & - & - & $14.3 \pm 4.4$ & - & - \\
11 & - & $10.5 \pm 3.3$ & - & - & - \\
13 & - & - & - & - & - \\
15 & $8.9 \pm 3.2$  & $17.7 \pm 4.4$ & - & $0.33 \pm 0.22$ & - \\
16 & $17.4 \pm 5.1$ & $17.3 \pm 5.4$ & - & $0.00 \pm 0.22$ & - \\ 
17 & $24.0 \pm 6.3$ & $22.5 \pm 6.8$ & - & $0.03 \pm 0.20$ & - \\ 
18/19 & $23.6 \pm 6.2$ & $44.5 \pm 8.6$ & - & $0.31 \pm 0.16$ & - \\
20 & $16.1 \pm 5.3$ & - & - & - & - \\
21 & $66.2 \pm 9.0$ & $264.2 \pm 16.7$ & $73.0 \pm 8.7 $ & $0.60 \pm 0.07$ &
$-0.57 \pm 0.06$ \\ 
22 & $39.6 \pm 7.2$ & $39.9 \pm 7.3$ & - & $0.00 \pm 0.14$ &  \\
24 & - & - & - & - & - \\
25 & $5.6 \pm 2.4$ & $7.8 \pm 2.8$ & - & $0.16 \pm 0.28$ & - \\
26 & $16.7 \pm 4.1$ & $22.8 \pm 4.9$ & $6.1 \pm 2.6$ & $0.15 \pm 0.16$ &
$-0.58 \pm 0.22$ \\
27 & - & $17.0 \pm 4.2$ & $21.7 \pm 4.8$ & - & $0.12 \pm 0.17$ \\
29 & - & $12.9 \pm 3.6$ & - & - & - \\ \hline
\end{tabular}
\end{center}
\end{table*}

\begin{figure}
\vspace{6.0cm}
\includegraphics{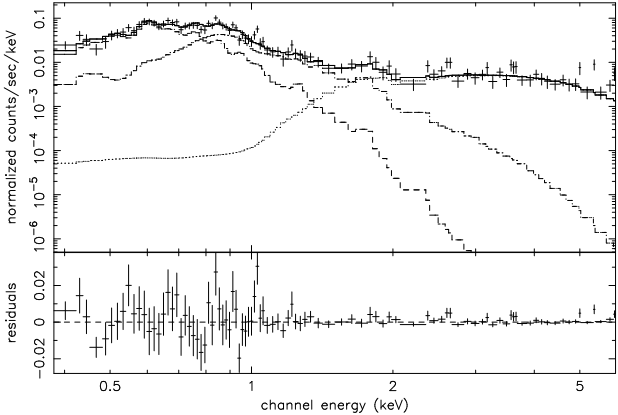}
\caption{Fitted spectrum of the NGC\,5253 diffuse emission as seen with
{\it Chandra}. The fit is an absorbed two temperature plus power-law
fit, with assumed Galactic column density of $N_{H(GAL)}=3.87 \times
10^{20}$~cm$^{-2}$. The fitted temperature components were $0.24 \pm
0.01$ and $0.75 \pm 0.05\kev$, with fitted absorbing column densities of
$(0.96 \pm^{0.13}_{0.12}) \times 10^{21}$ ~cm$^{-2}$ and $(1.96
\pm^{0.34}_{0.29}) \times 10^{21}$~cm$^{-2}$ respectively.  The
metallicity for the two thermal components were fitted simultaneously
giving a value of $0.14 \pm 0.01 Z_{\odot}$. The fitted absorbing column
density and photon index of the power-law component were $(3.74
\pm^{0.56}_{0.44}) \times 10^{22}$~cm$^{-2}$ and $\Gamma = 1.43
\pm^{0.06}_{0.07}$ respectively.}
\label{diffspec}
\end{figure}

\begin{table*}
\begin{center}
\caption{The gas parameters for the two thermal components of the diffuse
emission. The assumptions made are that the volume of the emitting
region is $V=1.54 \times 10^{64}$~cm$^{3}$ (assuming spherical
symmetry); $D=3.15$~Mpc; a filling factor, $f$; $n_e \sim(EI/Vf)^{1/2}$
where $EI$ is the emission integral (norm$\times 4\pi D^{2}$)/$10^{-14}$ and
norm is the normalisation obtained from the spectral fitting; $P\sim 
2n_e kT$,  $M\sim n_e m_{p}Vf$, $E_{Th} \sim 3n_e kTV$ and $t_{cool}
\sim (3kT)/(\Lambda n_e)$ where $\Lambda = L_{X}/EI$. Columns 2 and 3
give the values calculated from the {\it Chandra} data while columns 4 and
5 are the corresponding {\it XMM-Newton} values.}
\begin{tabular}{lccccc} \hline
&\multicolumn{2}{c}{\it Chandra}&\hspace*{0.5cm} & \multicolumn{2}{c}{\it
XMM-Newton}\\
\cline{2-3}\cline{5-6}
 & Soft component & Medium component & &Soft component & Medium component
\\ \hline 
$kT$~(\kev) & $0.24 \pm^{0.01}_{0.01}$ & $0.75 \pm^{0.05}_{0.05}$ && 
$0.25 \pm^{0.02}_{0.01}$ & $0.67 \pm^{0.04}_{0.05}$ \\
$T$ ($10^6$K) & 
$2.78\pm^{0.12}_{0.11}$ & $8.69 \pm^{0.58}_{0.58}$ && 
$2.90\pm^{0.11}_{0.12}$ & $7.76 \pm^{0.47}_{0.57}$  \\
$L_{X}$ ($10^{38}\ergs$) & 
$2.19\pm^{0.16}_{0.15}$ & $1.75\pm^{0.15}_{0.14}$ &&
$1.65\pm^{0.08}_{0.08}$ & $1.82\pm^{0.14}_{0.17}$ \\
$n_e$ [$\times f^{1/2}$] (cm$^{-3}$) & 
$0.065\pm^{0.002}_{0.002}$ & $0.041\pm^{0.002}_{0.001}$ && 
$0.060\pm^{0.001}_{0.001}$ & $0.044\pm^{0.002}_{0.002}$ \\
$E_{Th}$ [$\times f^{1/2}$] ($10^{54}$~erg) & 
$1.15\pm^{0.06}_{0.06}$  & $2.25 \pm^{0.18}_{0.18}$ &&
$1.11\pm^{0.04}_{0.06}$ & $2.21 \pm^{0.16}_{0.18}$ \\ 
$M$ [$\times f^{1/2}$] ($10^6\msun$)& 
$0.84\pm^{0.03}_{0.03}$ & $0.53\pm^{0.03}_{0.02}$ && 
$0.78\pm^{0.01}_{0.02}$ & $0.58\pm^{0.03}_{0.03}$\\
$P$ [$\times f^{-1/2}$] ($10^{-11}$ dyn~cm$^{-2}$) & 
$4.97\pm^{0.30}_{0.30}$ & $9.79\pm^{0.79}_{0.75}$ &&
$4.82\pm^{0.20}_{0.25}$ & $9.59\pm^{0.70}_{0.84}$ \\
$t_{cool}$ [$\times f^{1/2}$] ($10^8$yr) & 
$1.67\pm^{0.20}_{0.19}$ & $4.13\pm^{0.62}_{0.56}$ && 
$2.15\pm^{0.16}_{0.20}$ & $3.82\pm^{0.50}_{0.60}$ \\
$\Lambda$~($10^{-24}$\ergs~cm$^{3}$) & 
$3.36\pm^{0.34}_{0.33}$ & $6.78\pm^{0.84}_{0.77}$ &&
$2.95\pm^{0.19}_{0.22}$ & $5.95\pm^{0.65}_{0.78}$ \\
$EI$ ($10^{61}$cm$^{-3}$) & 
$6.52\pm^{0.45}_{0.47}$ & $2.58\pm^{0.23}_{0.21}$ &&
$5.59\pm^{0.23}_{0.31}$ & $3.06\pm^{0.24}_{0.28}$ \\ \hline
\end{tabular}
\end{center}
\end{table*}

\subsection{Diffuse Emission}

A background and point source subtracted spectrum of the region
containing the diffuse emission was extracted from the {\it Chandra}
data only, as the spatial resolution of {\it XMM-Newton} was inadequate
for resolving the point sources within the central region. This diffuse
emission spectrum is shown in Fig.~\ref{diffspec}. The fitted model
shown is an absorbed three component fit having two thermal components
and a power-law component -
$wabs_{GAL}(wabs(mekal)+wabs(mekal)+wabs(po))$. The fit gives $N_H=
(0.96 \pm^{0.13}_{0.12}) \times 10^{21}$~cm$^{-2}$, $(1.96
\pm^{0.34}_{0.29}) \times 10^{21}$~cm$^{-2}$ and
$(3.74\pm^{0.56}_{0.44}) \times 10^{22}$~cm$^{-2}$, for the fitted
absorbing column densities of the cool, warm and power-law components
respectively. The fitted temperatures are $kT \sim 0.24 \pm 0.01\kev$
for the soft component, and $kT \sim 0.75 \pm 0.05\kev$ for the medium
component, while the fitted photon index for the power-law component is
$\Gamma = 1.43 \pm^{0.06}_{0.07}$. The metallicities for the two thermal
components were fitted simultaneously, with $0.14 \pm 0.01 Z_{\odot}$.

The absorption corrected fluxes (and percentage contributions) in the
$0.37-6.0\kev$ energy band in each of these three components were: cool
thermal: $(1.84\pm^{0.13}_{0.13}) \times 10^{-13}\ergs$~cm$^{-2}$
($20\%$); warm thermal: $(1.47\pm^{0.13}_{0.12}) \times
10^{-13}\ergs$~cm$^{-2}$ ($15\%$) and power-law: $(6.48
\pm^{0.58}_{0.56}) \times 10^{-13}\ergs$~cm$^{-2}$ ($65\%$). The errors
shown here are the $90\%$ confidence levels for the normalisation value
obtained from the fit.  This result suggests that the diffuse emission
consists of at least two components at different temperatures and many
unresolved point sources.

As discussed in Section~3.1, several of the point sources in the central
region of NGC\,5253 could be young superbubbles and the subtraction of
such sources could therefore remove a substantial fraction of the
diffuse emission. Comparing the fluxes above with those obtained from
the analysis of the total emission gives the result that the point
source subtraction has removed $\sim 25\%$ of the cool thermal emission,
$\sim 60\%$ of the warmer thermal emission and $35\%$ of the emission
attributed to the power-law component, but these figures do not take
account of the fact that the fitted values for the absorbing column
densities have changed between the two fits. Re-fitting the diffuse
emission with the values fitted for the total emission and only allowing
the normalisations for the individual components to be free gives
reductions in the fluxes that correspond to $\sim 12\%$ for the cool
thermal emission, $\sim 46\%$ for the warmer thermal emission and $\sim
14\%$ for the emission from the power-law component. This does suggest
that the cooler thermal component is affected less by the point source
subtraction and so probably lies further away from the point sources and
star-clusters than the warmer component. When sources 16, 17, 18 and 20
are not subtracted from the total emission and this modified emission is
fitted as above with only the normalisations of each component as free
parameters, the reduction in the fluxes for each component are: cool
thermal: $\sim 10\%$; warm thermal: $\sim 43\%$; power-law: $\sim
13\%$. These figures suggest there is emission from all three components
of the fit associated with these four sources, with maybe slightly more
of the warmer thermal component than the other two, and yet the
temperatures measured for them during the spectral fitting was of the
order of that of the cooler component. As their contribution is not a
substantial fraction of the total emission in all three components,
there inclusion in the source subtraction does not effect the fluxes
measured for the diffuse emission greatly.

Other parameters of the two gas components have been calculated, from
the spectral fits to the {\it Chandra} data for the diffuse emission and
the {\it XMM-Newton} data for the total emission. The results are shown
in Table~7, including the effect of a filling factor $f$, and within
errors most of the calculated values agree between the two data
sets. The implications of these figures for the evolution of NGC\,5253
will be discussed further in Section~4. The assumption of spherical
symmetry and a filling factor of unity are both likely to be
overestimates, resulting in the quoted figures being underestimates for
$n_e$ and $P$ and overestimates for $M$, $E_{Th}$ and $t_{cool}$ (see
Strickland \& Stevens 2000 for a discussion of filling factors and their
likely values in galactic winds).

\begin{figure} \vspace{6.5cm}
\includegraphics{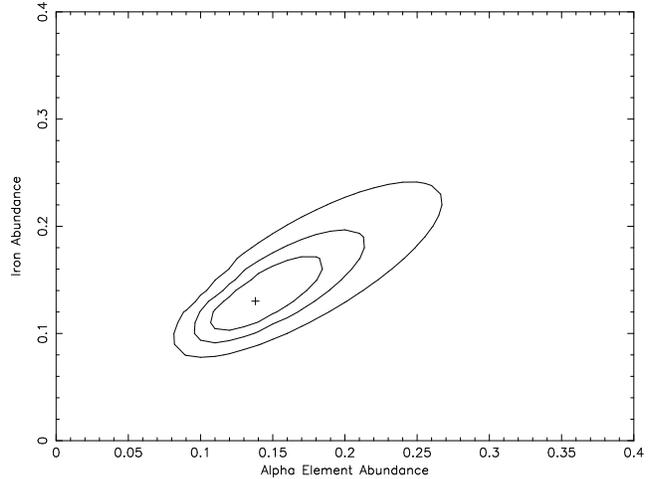} 
\caption{The confidence contours for the $\alpha$ and Fe element abundances
(see text for details). The units of both axes are the solar values for
the $\alpha$ elements and Fe respectively. The contours shown are the
68\%, 90\% and 99\% contours.}   
\label{alphaelem}
\end{figure}

Overall, the metallicity of the hot X-ray emitting gas is similar to
that obtained from optical spectroscopy of the central H{\small II}
regions (Webster \& Smith 1983) of the galaxy. Compared to other dwarf
galaxies though, the values of $0.14 \pm 0.01 Z_{\odot}$ from the
diffuse X-ray data and $0.19 \pm^{0.08}_{0.06} Z_\odot$ from the optical
data are low. Typical values for some other dwarfs are: Mrk~33: $\sim
0.3 Z_{\odot}$ (Legrand \etal\ 1997); NGC\,4449: $\sim 0.32 \pm 0.08
Z_\odot$ for the X-ray emitting gas (Summers \etal\ 2003); NGC\,1569: $>
0.25 Z_{\odot}$ for the X-ray emitting gas and a mean value of $0.2
Z_{\odot}$ for the ISM material (Martin \etal\ 2002). In a further
attempt to investigate the metallicities of the thermal components, and
in particular the ratio of the $\alpha$-elements to Fe, the spectrum was
re-fitted using variable abundance mekal models ({\it vmekal}) for the
two thermal components. The abundances of the two thermal components
were tied and the individual abundances of Mg, Ne, Si and Ca relative to
Solar values were tied to that of O to form the group of
$\alpha$-elements. The abundances of all the other elements were set to
the global value determined above.  The $\alpha/Fe$ ratio obtained in
this way was found to have no significant difference to the Solar value
at $(1.05 \pm 0.30) (\alpha/Fe)_{\odot}$.  The confidence contour plot
for these fits is shown in Fig.~\ref{alphaelem}, showing that the
$\alpha$ and Fe abundances are highly correlated, and that the fitted
ratio is $\sim 1$. We note that this is somewhat in contrast to the
values of Martin \etal\ (2002) for the dwarf starburst NGC\,1569, where
values of the $\alpha$ to Fe abundance ratio of $2.1- 3.9$ times the
Solar value are quoted
(and in the case of NGC\,1569 the photon statistics are rather better
than for NGC\,5253).  This value for NGC\,5253 is more in line with the
value of $(0.91 \pm^{0.07}_{0.10}) {(\alpha/Fe)}_{\odot}$ found for the dwarf
starburst NGC\,4449 (Summers \etal\ 2003) and may be indicative of
differences in the contributions made to the hot gas from SNe in these
different starbursts. The high values quoted for NGC\,1569 suggest the
presence of more Type II SNe, which would be expected if its current
starburst phase is $\sim 10 - 20$~Myr old as suggested in the literature
on this starburst (e.g., Israel \& De Bruyn 1988; Gonz\'{a}lez-Delgado
\etal\ 1997; Hunter \etal\ 2000). Such activity would give rise to more
$\alpha$-elements, as would a pattern of shorter more intense bursts of
star formation. The fact that NGC\,5253 has shown a fairly uniform rate
of star-formation throughout the galaxy over the last $10^{8}
-10^{9}$~yr (Caldwell \& Phillips 1989), would suggest that, with the
exception of the current burst in the very central region, which could
be very young and hence have little Type II SNe activity at present, its
star-formation and energy injection from star-forming regions has been
generally of a more continuous nature.

From Table~4, it can be seen that there is little temperature
variation evident within the sources, that appear to be extended (16,
17, 18, 19 and 20), in the central region of NGC\,5253, with the
exception of source 16 in the SW being cooler than the rest. The
absorbing column densities fitted to them are also fairly consistent and
there is overlap seen between adjacent sources within errors. The dust
lane in NGC\,5253 bisects the nuclear region such that sources 18 and 19
lie to the north of it and sources 16, 17 and 20 are to the south. From
the hardness ratios and source counts in various energy bands shown in
Tables~5 and 6, the two northern sources are
harder than those in the south, suggesting that they should be either
hotter or lie behind larger absorbing columns. Source 17's temperature
and 16's absorbing column density would argue against this. The latter
could be explained by the western side of the galaxy being tilted away
from us. The low count-rate statistics however, hamper the drawing of
any firm conclusions, though the results do suggest that these objects
are similar in nature to each other.

\begin{figure}
\vspace{7.5cm}
\includegraphics{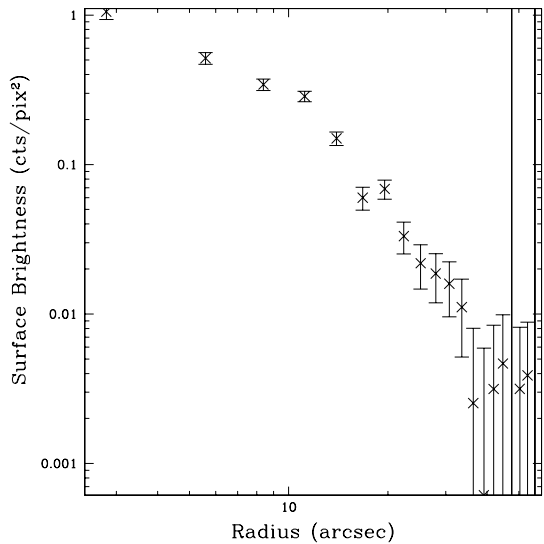}
\caption{Background subtracted radial surface brightness plot, showing the
extent of the diffuse X-ray emission in the $0.3-8.0\kev$ band. The
counts were extracted in 20 concentric annuli, each with a width of
$2.75^{''}$ centred on $\alpha = 13^{h} 39^{m} 56.2^{s}$ and $\delta =
-31^{\circ} 38{'} 29.9{''}$. The diffuse emission extends out to 
$\sim 35^{''}$ ($\sim 0.53$~kpc) by which point, the emission has
dropped to the background level.}
\label{rad}
\end{figure}

Also, from the 3 colour {\it Chandra} image, further differences can be
seen in the nature of the point sources in these three energy bands.
Sources 17, 18, and 20 all appear very red on the three colour image,
reinforcing their similar nature. Sources 16 and 19 show slightly
different behaviour, by having more emission in the medium band than the
other three sources, which is reflected in their yellow colour on
Fig.~\ref{3col}. The diffuse emission is mainly confined to the central
region of the galaxy, and the peak of its emission is centred on the
region occupied by sources 17, 18 and 19, which also corresponds to the
region of peak intensity in the H$\alpha$ emission. The images of the
{\it XMM-Newton} data shown in Figs.~\ref{mosaic} and~\ref{xmmchan} show
that the diffuse X-ray emission extends away from the central region in
several directions, although as stated earlier most of the extension to
the SW appears to be associated with point sources. This is confirmed by
the log-log plot of the radial surface brightness profile shown in
Fig.~\ref{rad}, produced by extracting the counts from 20 concentric
annuli each with a width of $2.75{''}$ centred on a position coincident
with the centre of the $D_{25}$ ellipse at $\alpha = 13^{h} 39^{m}
56.2^{s}$ and $\delta = -31^{\circ} 38{'} 29.9{''}$. This figure shows
that the diffuse emission can only be traced out to a distance of $\sim
35{''}$ (0.53~kpc for our assumed distance) before it falls to a level
comparable to that of random background fluctuations. The preferred
directions for the observed extensions in the diffuse emission seem to
be along the minor axis of the galaxy, both to the NW and SE. If these
extension in the diffuse X-ray emission are due to superbubbles and
superwinds then their activity should have swept-up shells and produced
filaments which should be evident in the H$\alpha$ image of the
galaxy. To look for such evidence, of correlations between the extended
X-ray emission and structure within the H$\alpha$ emission, X-ray
contours from the point source and background subtracted diffuse
emission have been overlaid on the H$\alpha$ image of the central region
of NGC\,5253. In addition, the detected sources in the central region of
the galaxy (sources 15--22) are also overlaid on the image to show the
relative positions, and extent of these sources (as determined by the
{\it wavdetect} program), compared with the
regions of increased star-formation on the H$\alpha$ image. The results
are shown in Fig.~\ref{hax} and the correlations between the various
components are discussed further in Section~4.1.1. As a further
comparison between emission at different wavebands, Fig.~\ref{opt} shows
the positions of the six largest star clusters found in the central
region of NGC\,5253 overlaid on the {\it Chandra} soft ($0.3- 1.1\kev$)
and medium ($1.1-3.8\kev$) smoothed images of the galaxy along with the
positions of the X-ray sources 15--22 as shown in Fig.~\ref{hax}. These
images show only two strong correlations between the most northern and
southern star clusters shown here and X-ray sources. Most of the soft
X-ray emission occurs between and around these large star clusters in
regions where many smaller clusters are to be found. The second most
northern of these large star clusters shown on these images has four
smaller clusters in close proximity to it, yet it appears to have little
X-ray emission close to it. However, these clusters are located at the
western end of the dust lane, which is most likely responsible for the
absorption of the X-ray emission seen both in these images and on
Fig.~\ref{3col}.

\section{Discussion}

\begin{figure*}
\vspace{13cm}
\includegraphics{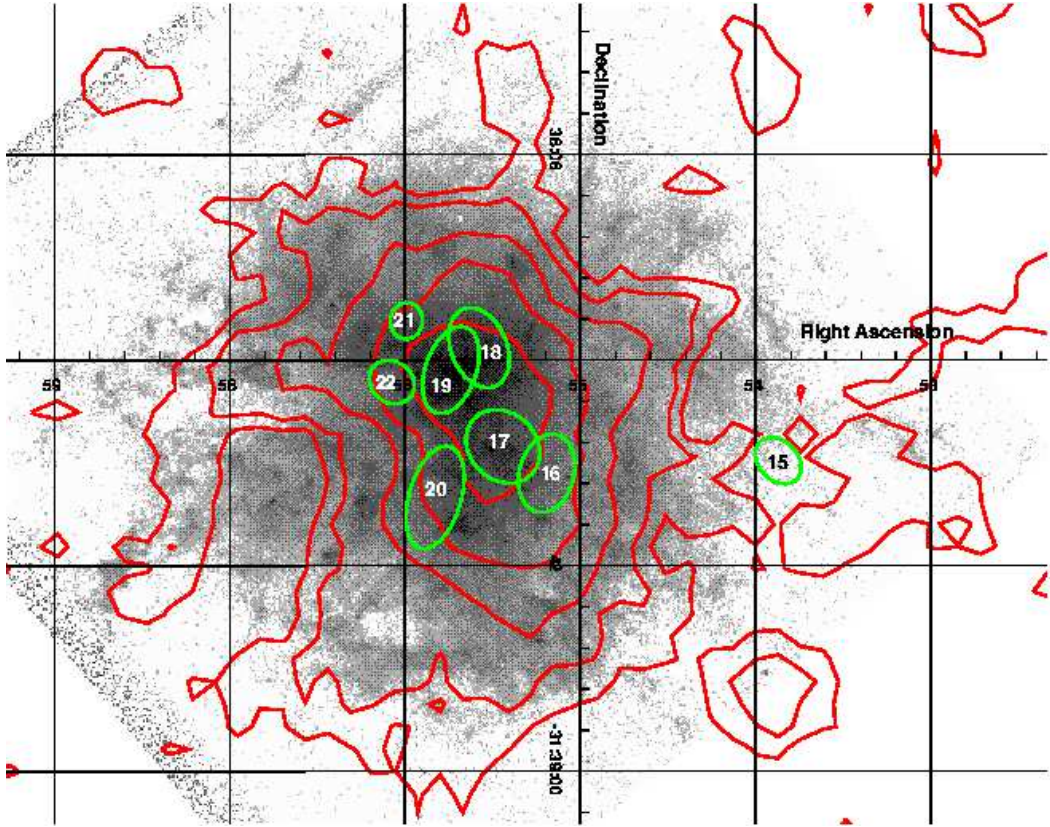}
\caption{X-ray contours from the {\it Chandra} diffuse emission
($0.3-8.0\kev$ energy band) overlaid on an H$\alpha$ image of
NGC\,5253. The X-ray contours shown are at flux densities of $0.26
\times 10^{-13}$, $0.52 \times 10^{-13}$, $1.04 \times 10^{-13}$, $2.08
\times 10^{-13}$ and $4.16 \times
10^{-13}\ergs$~cm$^{-2}$~arcmin$^{-2}$. Also shown are the positions and
extent of the sources detected in the central region of the {\it
Chandra} data and these are labelled with their ID numbers as shown in
Fig.~\ref{sources}. North is to the top and East to the left.} 
\label{hax}
\end{figure*}

\begin{figure*}
\vspace{8cm}
\includegraphics{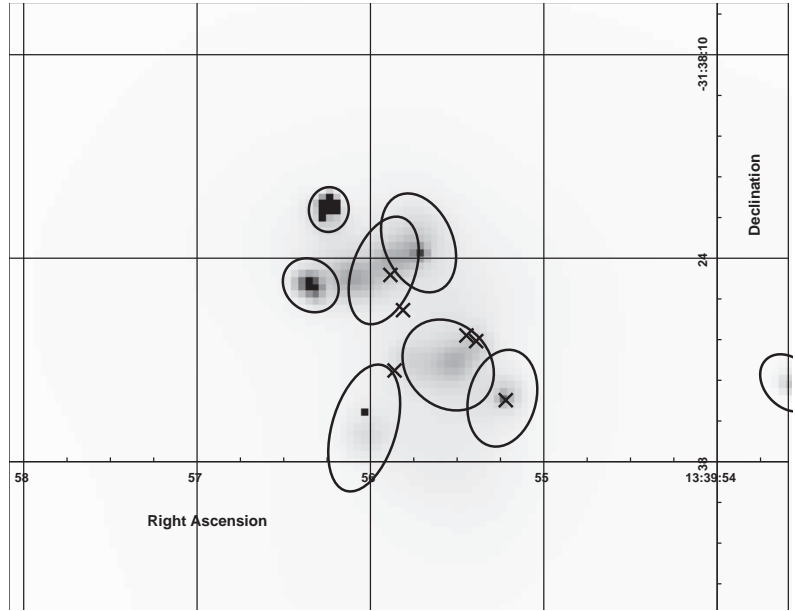}
\caption{Positions of the six largest optical star clusters are shown as
crosses and these are overlaid on the {\it Chandra} soft-band ($0.3 -
1.1\kev$) smoothed X-ray images of
NGC\,5253. Also overlaid on these images are the X-ray sources detected
in the central region of the {\it Chandra} data, which are labelled in
Fig.~\ref{hax}. N is to the top and E to the left and the image is
a linear greyscale. The flux densities range from 0 to $7.0 \times
10^{-11}\ergs$~cm$^{-2}$~arcmin$^{-2}$.}
\label{opt}
\end{figure*}

It is apparent from Fig.~\ref{hax} that both the diffuse X-ray emission
and the H$\alpha$ emission are mainly confined to a region of radius
$\sim 0.5$~kpc in the centre of NGC\,5253. The most intense X-ray
emission is coincident with the most intense emission seen in the
H$\alpha$ image in the Northern half of the nucleus, suggesting that
this emission is associated with the increased star-formation occurring
in these regions. In general, higher column densities were fitted for
the sources to the W and S of the galaxy, as seen in Table~4,
and would suggest that these areas lie further away from us behind more
absorbing material than the N and E of the galaxy. This would mean the
minor axis lies away from our line-of-sight (tilted both to the N and
W), and that the galaxy is also rotated about this axis in such a way
that the Southern side of the galaxy disk is tilted away from us,
consistent with NGC\,5253 being a dwarf elliptical galaxy that is
inclined to our line-of-sight. The bulging of the diffuse emission seen
to the NW and SE in the X-ray contours and {\it XMM-Newton} images,
could be indicative of a bipolar outflow along the minor axis of the
galaxy, as would be expected from standard superbubble models (Weaver
\etal\ 1977).

\subsection{Comparison with Observations at Other Wavelengths}

\subsubsection{H$\alpha$ Emission}

NGC\,5253 contains many regions of increased star-formation within the
central region of the galaxy, which are clearly visible in H$\alpha$
observations, with the most intense regions being seen in the North of
the nucleus, coincident with the end of the dust lane and weak CO
emission (Turner \etal\ 1997). This is the region that contains two of
the brightest star-clusters analysed by Tremonti \etal\ (2001) and
corresponding to our detected X-ray sources 18 and 19. The extension
seen in the diffuse emission to the WNW of this region could be the
result of an outflow induced by the stellar winds of the massive stars
present in these star clusters. This region also lies close to the
position of the `radio super-nebula' discussed by Gorjian \etal\ (2001),
which they have interpreted as an highly obscured, very young globular
cluster, rather than an obscured AGN (see also Mohan, Anantharamaiah \&
Goss 2001; Turner \etal\ 2003).

As well as many of the star clusters coinciding with X-ray sources, some of
the filamentary and shell-like structures seen in H$\alpha $ images seem to
have counterparts in the diffuse X-ray emission. The small H$\alpha$ shell
seen to the SSE in Fig.~\ref{hax} appears to be filled with diffuse X-ray
emission and the radial filament extending off the H$\alpha$ image to the
ESE is also associated with what appears to be a filamentary extension of
the diffuse X-ray emission. The field-of-view of the H$\alpha$ image shown
in Fig.~\ref{hax} is too small to show the $\sim 1$~kpc shells detected to
the West of the galaxy by Marlowe \etal\ (1995), but these can be seen in
Fig.~2j of Martin (1998). The orientation of the more northern of these two
shells is roughly aligned with the galaxies minor axis. The slow expansion
speed of these shells [$\sim 35\kms$ (Marlowe \etal\ 1995)] and the
fact that the diffuse X-ray emission does not extend out to them, as shown
in the radial surface brightness plot of Fig.~\ref{rad}, may mean that
these shells are not associated with the current burst of star formation
but are remnants of earlier starburst activity. The presence of more than
one bubble having been blown close to the minor-axis on the West side
suggests that what is being seen in this galaxy is much more complex than
the simple bipolar outflow scenario assumed to explain the effect of
starburst activity. This phenomenon could be the result from a young
starburst, in so much as the outflows from individual star-clusters may not 
have overlapped or merged to produce what would be observed as a single
outflow. The presence of the dust lane bisecting the nuclear region may
also be acting in such a way as to channel any superbubbles so that they
expand preferentially into areas where the density of the ambient ISM is
less. The possible dynamics of such superbubbles will be discussed later.

\subsubsection{Infrared and Radio Emission}

Strong mid-infrared emission (Gorjian \etal\ 2001) is detected
from a source in the northern half of the central region of the galaxy just
at the Northern edge of our detected sources 18 and 19. This is an
extremely compact source occupying a region with diameter $\sim 1-2$~pc
and needing to contain $\sim 10^{5}-10^{6}$ stars to supply its ionization
requirements. This object is most likely a very young (no more than a few
$\times 10^{6}$ years old) globular cluster.

Just to the south of this infrared emission, at a position corresponding
to our source 19, is the peak of the 2cm radio emission (Beck \etal\
1996). While mainly concentrated in the Northern half of the central
region of the galaxy, some 2cm emission is also detected at the
positions of our sources 17 and 20 but not coincident with source
16. This radio emission has a flat spectrum, with a spectral index of
$-0.1 \pm 0.01$, which is attributed to thermal bremsstrahlung emission
from the H{\small II} in the region.  Relatively low levels of
non-thermal synchrotron emission are seen in NGC\,5253 (Turner \etal\
1998), which is unusual for a starburst galaxy and again suggests
that the starburst occurring in the central region of NGC\,5253 could be
young.

As mentioned earlier, VLA 21~cm observations of NGC\,5253 have suggested
that the bulk of the {\small HI} rotates about the galaxy's major axis
(though see Meier \etal\ 2002 for evidence of infall into NGC\,5253).
Fig.~1 of Kobulnicky \& Skillman (1995), shows that the neutral gas
extends to about twice the optical extent of the galaxy along its minor
axis, occupying a region of $\sim 4{'}$ in diameter. This is well beyond
the extent of the diffuse X-ray emission and H$\alpha$ emission seen in
the central region. The peculiar rotation could result from large-scale
outflow along the minor axis, but such behaviour should give rise to a
shell-like centrally evacuated structure in H{\small I} emission, as
seen in the $\sim 1$~kpc H$\alpha$ shells. The column density measured
for the H{\small I} increases towards the centre and peaks at a position
close to our detected source 20. This is offset by about $15{''}$ from
the peaks of emission seen at shorter wavelengths and is in the southern
rather than northern half of the galaxy, however this does coincide with
the point where the dust lane enters the nuclear region from the ESE.

\subsection{Potential for Blow-out and its Implications on the Fate of
NGC\,5253}

Although at present there does not appear to be a clearly defined single
superbubble structure within NGC\,5253, the presence of hot,
over-pressured gas with a long cooling time, should lead to an outflow
from the central region. With this in mind, in this section we apply the
standard superbubble model to the hot gas of NGC\,5253 to investigate
the potential of the galaxy to deposit energy, metal-enriched material
and ISM material into the intergalactic medium.

The age of the current burst of star-formation in the centre of
NGC\,5253 is not well determined, and probably does not have a single
age. The comparative lack of a non-thermal signature in the
radio-spectrum (Beck \etal\ 1996; Turner \etal\ 1998) may suggest a lack
of supernova activity and the detected presence of Wolf-Rayet stars
(Schaerer \etal\ 1997) in regions corresponding to our detected source
19 suggest that star-formation has been occurring for only a few Myr in
this region. Typically ages for star clusters in NGC\,5253 reported in
the literature are: 7.9~Myr (Marlowe \etal\ 1999); 2.8~Myr and 4.4~Myr
for two separate regions within the nucleus (Schaerer \etal\ 1997);
2~Myr for the brightest star cluster (Tremonti \etal\ 2001); 5~Myr
(Martin \& Kennicutt 1995); 3~Myr (Mas-Hesse \& Kunth 1999), though as
noted earlier there are some older clusters (Calzetti \etal\ 1997), and
indeed these older star-clusters can still be contributing mechanical
luminosity.

Based on these figures we adopt an age of 5~Myr for the current
starburst, in order to perform some analysis of the dynamics of the
probable superbubbles in NGC\,5253. We recognise the limitation of this
approach in such a complex system as NGC\,5253, where the
star-formation is not co-eval. The northern star-clusters in NGC\,5253
are younger and here the analysis may be useful.

The bulk of the diffuse X-ray emission is confined within a radius of
$\sim 0.6{'}$ (0.53~kpc for our assumed distance), that corresponds to the
same region containing the majority of the H$\alpha$ emission as shown
in Fig.~\ref{hax}. For simplicity this will be treated as a single
superbubble resulting from the combined action of the multiple
star-clusters within the starburst. The dynamic time required for the
two $\sim 1$~kpc H$\alpha$ shells detected by Marlowe \etal\ (1995) to
have evolved is $\sim 10$~Myr, assuming they have had a constant expansion
velocity of $35\kms$ (Marlowe \etal\ 1995) for this time. This value is
greater than the estimated duration of the current burst of
star-formation, and the diffuse emission does not appear to extend this
far out. However, this figure is also likely to be an over estimate
since their speed will have been greater in the past, which would bring
their expansion times closer to the higher age estimates of the star
clusters. These observations raise the question of what is responsible
for their existence which will be discussed further later.

The combined action of the stellar winds of the massive O and B stars in
the star-forming regions in the centre of NGC\,5253 will blow a
superbubble that can be modelled using the starburst driven outflow
models of Castor, McCray \& Weaver (1975) and Weaver \etal\ (1977). In
such models, the stellar winds from the OB associations sweep up and
shock a thin, dense H$\alpha$ emitting shell of ISM material, with a
contact discontinuity just behind the shell, separating the shocked wind
material from the shocked ISM material. The innermost region of the
bubble will contain the freely expanding supersonic wind bounded by an
inward-facing shock. In the `snow-plough' phase, the shocked wind
material occupies most of the bubbles interior and most of the bubble's
mass is contained in the H$\alpha$ shell. The expansion of the bubble is
driven by the over-pressured interior, which, as can be seen from the
figures of Table~7 is of an order of magnitude greater than
typical values for the ambient ISM within the Milky Way which is $\sim
10^{-12}$~dyn~cm$^{-2}$. The radius $R_{B}$ and expansion velocity
$v_{B}$, of the bubble during this phase, are given by

\begin{equation}
R_{B} = 168 \left (\frac {L_{mech,40}}{n_0} \right )^{1/5} 
t_{6}^{3/5}\;\;\;{\rm pc} 
\end{equation}

\begin{equation}
v_{B} = 99 \left(\frac {L_{mech,40}}{n_0}\right)^{1/5}
t_{6}^{-2/5}\;\;\;{\kms}{\rm ,}
\end{equation}

\noindent where $L_{mech,40}$ is the mechanical luminosity injected by
the starburst in units of $10^{40}\ergs$, $t_6$ is the age of the
starburst in units of $10^{6}$~yr and $n_0$ is the ambient atomic number
density of the ISM.

The mechanical luminosity ($L_{mech}$) injected by the starburst can be
estimated from the stellar populations in the starburst. Marlowe \etal\
(1999) have performed a detailed analysis of the starburst in
NGC\,5253, using starburst evolutionary synthesis models (c.f. Leitherer
\& Heckman 1995; Leitherer \etal\ 1999) and they estimate that the
energy injection rate in NGC\,5253 is $L_{mech}=3.6\times 10^{40}\ergs$
for both instantaneous burst and continuous star-formation models, and we 
shall adopt this figure hereafter. The mass injection rate from the {\sl
Starburst99} models for this level of mechanical luminosity will be
$\sim 0.02-0.04\msunyr$ depending on whether instantaneous or
continuous starburst models are assumed (for an age of 5~Myr).

An estimate of the ambient density in the central regions, $n_0$, can be
made based on the H{\small I} observations of Kobulnicky \& Skillman
(1995). The peak H{\small I} column density in the central regions is
$2.6\times 10^{21}$~cm$^{-2}$ which, assuming that the path length is
$\sim 1$kpc (i.e comparable to the beam width) leads to
$n_0=1$~cm$^{-3}$.  We note that Martin (1997) inferred densities of
$\sim 250$~cm$^{-3}$ in the warm photoionised gas in the core of
NGC\,5253, though with a filling factor of a percent or so, and that
much higher densities are also implied for the smaller volumes
associated with the H{\small II} regions in NGC\,5253 (Turner \etal\
1998).

Using these figures in equations 1 and 2 give values for $R_{B}$ and
$v_{B}$ of 570~pc and $67\kms$. The size estimate is comparable to the
size of the region occupied by the bulk of the X-ray and H$\alpha$
emitting gases, but no allowance has been made for the effect of the
interaction of outflows from the multiple star-clusters present in the
central region, while the theory is applicable to the growth of a single
bubble. This size estimate is smaller than the size of the reported
large-scale H$\alpha$ shells, their expansion velocities were measured
to be $\sim 35\kms$, which is of roughly the same order as predicted
here. It is possible that we may be observing the escape of some stellar
wind material through ruptures in the original bubble that are then
proceeding to sweep-up and shock more of the ISM lying outside the
central region. The emission from such gas has to be at a very low
intensity though as it is not detected by either {\it Chandra} or {\it
XMM-Newton}. The average count rate in the annuli beyond the 0.53~kpc to
which the X-ray emission can be traced is $0.05 \pm 0.28 \times
10^{-3}$~cts~s$^{-1}$, which corresponds to a flux of $\sim 0.3 \pm 1.9
\times 10^{-15}\ergs$~cm$^{-2}$.  Alternatively, a clumpy environment
could allow preferential escape along channels between molecular clouds,
although there does not appear to be evidence for the existence of such
objects, and the low expansion velocities observed would argue against
such uninhibited outflow. There are however two `dark clouds' reported
in the West of the galaxy by Hunter (1982).

The superbubble model allows prediction of the X-ray luminosity
of an expanding bubble to be made. From Table~7, it can be
seen that the radiative cooling time for both components of the X-ray
emitting gas is long compared to our assumed expansion time of
5~Myr, and so radiative losses can be assumed to be small. Assuming 
spherical symmetry for the expansion, uniform density for the
ambient ISM and a constant rate of kinetic energy injection, the X-ray
luminosity is given by

\begin{equation} 
L_{X} = \int n(r)^{2}\Lambda_{X}(T,Z)dV
\end{equation}

\noindent where $\Lambda_{X}(T,Z)$ is the volume emissivity of the gas
being considered. Based on the density $n(r)$ and temperature $T(r)$
profiles (Chu \& Mac Low 1990), this equation can be evaluated to yield
the X-ray luminosity of the superbubble. Assuming an emissivity of the
X-ray emitting gas of $\Lambda=6.8\times 10^{-24}\ergs$~cm$^{3}$ (cf
Table~7), we find that for NGC\,5253

\begin{equation} 
L_{X} = 9.3 \times 10^{36} L_{mech,40}^{33/35} n_{0}^{17/35} t_{6}^{19/35}
\;\;\;{\ergs}\,,
\end{equation}

As discussed earlier, assuming an energy injection rate from the
starburst in NGC\,5253 of  $L_{mech}=3.6\times 10^{40}\ergs$, an ambient 
density of $n_0=1$~cm$^{-3}$ and an age of 5~Myr, leads to a predicted 
thermal X-ray luminosity for NGC\,5253 of $L_X=0.7 \times 10^{38}\ergs$,
which is lower than that observed for the two thermal components
($L_x\sim 4\times 10^{38}\ergs$, but given the uncertainties in the
some of the assumed parameters, and indeed the validity of treating the 
situation in NGC\,5253 as a single superbubble, this is perhaps
reasonable agreement.

We note that an estimate of the mechanical energy injection rate can
also be obtained from the thermal energy contained within the diffuse
X-ray emission. Averaging the values obtained from the {\it Chandra} and
{\it XMM-Newton} data, the total thermal energy is $3.4 \times
10^{54}$~erg. Assuming this is the result of complete thermalization of
the stellar winds for the duration of the current starburst, then
$L_{mech}$ is $\sim 2.2 \times 10^{40}\ergs$, which is again in broad
agreement with that from theoretical considerations of the starburst
stellar population.

The total mass of material contained in the X-ray emitting gas obtained
from the average figures of the two data sets is $\sim 1.4 \times
10^{6}\msun$ (ignoring the filling factor). If this mass has been
injected during the lifetime of the starburst then the mass injection
rate is on average $\sim 0.3 \msunyr$. 

Another estimate of the mass injection rate can be obtained from the
fact that for an adiabatic superbubble, the X-ray temperature $T_{X}
\sim 5/11T_{0}$, where

\begin{equation}
T_{0} = \frac {2}{3} \left ( \frac {L_{mech}}{\dot M} \right ) \frac
{\mu m_{H}}{k}\;\;\; {\rm K}
\end{equation}

\noindent (Heckman \etal\ 1995) and is the temperature attained if all
the mechanical energy injected by the starburst remains entirely within
the shocked wind region. The weighted average temperature of the diffuse
X-ray emitting gas (Table~7) is $\sim 5
\times 10^{6}$~K, which gives a value for $T_{0} \sim 1.1 \times
10^{7}$~K and from the assumed $L_{mech}$, a mass injection rate of
$\dot M \sim 0.25\msunyr$. Finally, an estimate of the deposited mass
from both the stellar winds and conductive evaporation of the swept-up
ISM material can be made from the density of the diffuse X-ray emitting
gas. This figure depends on the filling factor of the bubble interior
but assuming $f=1$ results in an upper limit. Further, it is necessary
to assume that any volume occupied by the swept-up cool shell of ISM and
the freely expanding stellar winds is small in comparison to the shocked
wind region.  With the combined density of the soft and medium
components, $n \sim 0.1$~cm$^{-3}$, $V=1.54 \times 10^{64}$~cm$^{3}$ and
all other symbols having their usual meanings,

\begin{equation} 
M = \rho V = n \mu m_{H} \frac {4}{3} \pi R^{3} \sim 8\times
10^{5}\msun \,,
\end{equation} 

\noindent which implies a mass injection rate of $0.16\msunyr$ over 5~Myr. 

The mass injection rates into the hot phase derived here from the X-ray
observations are thus either slightly above or broadly comparable to the
current star-formation rate ($0.2\msunyr$) derived from infrared and
H$\alpha$ luminosities (Table~1), but rather higher than
the current stellar mass-injection rates.

The mass deposition rate into the hot-phase seen here suggests that if
the outflow from the starburst region develops into a superwind and
escapes the gravitational potential well of the galaxy then there is
substantial scope for both metal-enrichment of, and deposition of large
amounts of mass and energy into the IGM. The likelihood of blow-out
occurring can be assessed by applying the criterion for blow-out,
defined by Mac Low \& McCray (1988) to NGC\,5253. This defines a
parameter $\Lambda$, the dimensionless rate of kinetic energy injection,
and predicts blow-out if $\Lambda\geq 100$, with $\Lambda$ given by

\begin{equation}
\Lambda = 10^{3} L_{mech, 40} H_{kpc}^{-2} P_{4}^{-3/2} n_{0}^{1/2}
\end{equation}

\noindent where $L_{mech, 40}$ is the mechanical energy luminosity in
units of $10^{40}\ergs$, $H_{kpc}$ is the galaxy scale-height in
kpc, and $P_{4}$ is the initial pressure of the ISM in units of $P/k =
10^{4}$~K~cm$^{-3}$. For $\Lambda \geq 100$, with $L_{mech, 40} = 3.6$,
$n_{0}=5$~cm$^{-3}$ and assuming $P_{4} \sim 1$ (typical of the
value in the Milky Way), then $H_{kpc} \leq 9$. The H{\small I} halo
of NGC\,5253 extends to around twice the optical extent $\sim 2'$
($\sim 1.83$~kpc) along the minor axis of the galaxy (Kobulnicky \&
Skillman 1995). It seems extremely likely then that the hot,
metal-enriched gas generated by the current starburst along with the
swept-up and conductively evaporated ISM material can escape the
gravitational potential well of NGC\,5253.

\section{Summary and Conclusions}

In summary, we have presented an X-ray analysis of the important dwarf
elliptical, starburst galaxy NGC\,5253 using data from both the {\it
Chandra} and {\it XMM-Newton} satellites. We detected X-ray emission
from 31 discrete point sources in the {\it Chandra} ACIS-S3 chip data
with 17 of these lying within the optical extent of the galaxy, as
measured by the $D_{25}$ ellipse. Some of these can be clearly
identified as X-ray binaries while others seem to be associated with the
emission from super star-clusters rather than individual point sources,
as they are extended compared to the {\it Chandra} PSF. The 5 sources in
this category (sources $16 - 20$) all lie in the central region of the
galaxy and are associated (particularly in the case of sources 17, 18
and 19) with the most intense regions of emission seen on the H$\alpha$
image. The unabsorbed luminosities of these sources in the $0.3 -
8.0\kev$ energy band, range from $(0.25 - 1.50) \times 10^{38}\ergs$.

The majority of the X-ray emission is confined to the central region of
the galaxy where the intense, current burst of star-formation is
occurring. The extent of the emission also correlates well with that of
the H$\alpha$ emission and both are confined to a region of diameter
$\sim 1$~kpc in the centre of the galaxy. Both of these emissions are
embedded within a spherical distribution of H{\small I} that extends out
to $\sim 2$~kpc from the centre.

The diffuse X-ray emission in NGC\,5253 requires two thermal and one
power-law component to best fit its spectrum, suggesting that the hot
gas is a multi-phase environment with unresolved point sources, probably
lower luminosity X-ray binaries, embedded in it. The fitted gas
temperatures are $0.24 \pm 0.01\kev$ and $0.75 \pm 0.05\kev$ for the
soft and medium components respectively and their respective absorption
corrected luminosities in the $0.37 - 6.0\kev$ band are 
$(2.19 \pm^{0.16}_{0.15}) \times
10^{38}\ergs$ and $(1.75 \pm^{0.15}_{0.14}) \times 10^{38}\ergs$. This
emission is seen to be extended in several directions in the {\it
Chandra} data but this is not confirmed by the {\it XMM-Newton}
data. However, the extent is not as great as the $\sim 1$~kpc shells
reported to the NW and WSW by Marlowe \etal\ (1995), in the H$\alpha$
emission. These shells are expanding with velocities of $\sim 35\kms$
and hence have required $\sim$10~Myr to have evolved. NGC\,5253 is
proving to be a rather enigmatic object though, as our analysis of the
dynamics of the hot gas are not consistent with the expansion velocity
and age of these H$\alpha$ shells.

The hot X-ray emitting gas has a total thermal energy content of $3.4
\times 10^{54}$~erg and a total mass of $1.4 \times 10^{6}\msun$. From
values in the literature, we assume an age for the starburst responsible
for this hot gas of 5~Myr and a value for the density of the ambient
ISM of $n_{0} = 5$~cm$^{-3}$. The radius of the region in which most
of the diffuse X-ray and H$\alpha$ emission is confined is
$0.53$~kpc. Application of standard superbubble models to NGC\,5253's
starburst, assuming the age and density above, result in a predicted
radius and expansion velocity for a single superbubble of $\sim 400$~pc
and $\sim 50\kms$, figures which are comparable to those observed. 

The main
problem here would seem to lie in the fact that no clear bipolar
outflow is observed in NGC\,5253 and no allowance has been made for the
presence of multiple superbubbles in the central region. If the central
starburst is young, the bubbles blown around individual star clusters
will overlap and impede each others expansion confining the emission to
a smaller region than expected and reducing the observed expansion
rate. The H$\alpha$ shells, reported by Marlowe \etal\ (1995), to the
West could be the result of the superbubbles closest to the edge of the
central region being able to expand more freely. The lower velocity seen
could result from less stars contributing to the energy input in these
regions and/or the outflow having experienced some recent deceleration
on encountering higher density material such as the dark clouds reported
in the West of the galaxy (Hunter 1982). However, our radial surface
brightness profile does not show any detectable X-ray emission beyond
$\sim 0.5$~kpc.

NGC\,5253 is a very complex object, which shows that standard
superbubble models are difficult to apply to dwarf starbursts
galaxies. However, the energy injection rate into the galaxy would seem
sufficient to allow the expanding hot gas to escape the gravitational
potential well of NGC\,5253 and its relatively small ($\sim 1.8$~kpc
along the minor axis) H{\small I} halo would also offer little
resistance to this. Thus, it seems likely then that NGC\,5253 will lose
metal-enriched material, mass and energy as a result of its current bout
of star-formation.

\section*{Acknowledgements}
The referee is thanked for a helpful report. 
LKS and IRS acknowledge funding from a PPARC studentship and Advanced
Fellowship respectively. DKS is supported by NASA through {\it Chandra}
Postdoctoral Fellowship Award Number PF0-10012, issued by the {\it Chandra}
X-ray Observatory Center, which is operated by the Smithsonian
Astrophysical Observatory for and on behalf of NASA under contract
NAS8-39073. Our thanks go to Christy A. Tremonti for kindly providing us with
the H$\alpha$ image of NGC\,5253.

\end{document}